\documentclass[aps,pra,10pt,twocolumn,nofootinbib]{revtex4-2}%
\pdfoutput=1%
\usepackage{graphicx}%
\graphicspath{{img/}}%
\usepackage{amsmath, amssymb}%
\usepackage{mathphyscmd}%
\setcitestyle{super}%
\newcommand\papertitle{How Shaped Waves Propagate in Scattering Media}%

\usepackage{hyperref}%
\hypersetup{%
	pdfinfo={%
		Author       = {David Gaspard and Arthur Goetschy},
		Title        = {\papertitle},
		Subject      = {General Physics},
		CreationDate = {D:20250910173825+0200}, %
	},
	pdfborderstyle={/S/U/W 1}, %
	breaklinks=true
}

\begin{document}%
\title{\papertitle}%

\author{David Gaspard}%
\email[E-mail:~]{david.gaspard@espci.psl.eu}

\author{Arthur Goetschy}%
\email[E-mail:~]{arthur.goetschy@espci.psl.eu}

\affiliation{\href{https://ror.org/00kr24y60}{Institut Langevin}, \href{https://ror.org/03zx86w41}{ESPCI Paris}, \href{https://ror.org/013cjyk83}{PSL University}, \href{https://ror.org/02feahw73}{CNRS}, 75005 Paris, France}
\date{\today}

\begin{abstract}%
The radiative transport equation provides a powerful framework for describing wave propagation in scattering media.
However, it cannot describe coherent waves whose incident wavefront is deliberately tailored.
Here we establish a transport theory for wavefront-controlled waves based on a matrix transport equation.
Unlike conventional radiative transport, which propagates a scalar radiance, our theory propagates a complex two-by-two matrix that retains phase information and obeys a nonlinear transport equation, despite the underlying wave dynamics being linear.
We use this matrix transport equation to derive the spatial profiles of the energy and current densities of transmission eigenchannels, from closed to open channels, in arbitrary diffusive systems.
Remarkably, these profiles can be expressed analytically in terms of the solution of the conventional radiative transport equation for a random wave in the same medium.
The theory further extends to energy loss, including absorption and leakage into uncontrolled channels, revealing symmetry breaking of transmission eigenchannels in complex structures.
Validated against numerical solutions of the wave equation, we establish matrix transport as a framework for describing and controlling coherent wave propagation in complex media.
\end{abstract}%
\keywords{Suggested keywords}%
\maketitle

\par One of the most powerful tools for describing light propagation in turbid media---such as the atmosphere, materials, or biological tissues---is the radiative transport equation (RTE).
This equation describes the propagation of radiance, an intensity that depends on position and direction, which can be interpreted as a statistical distribution of rays.
The RTE has the same structure as the linear Boltzmann equation used in statistical and quantum physics for electrons in solids, ions in plasmas, or neutrons in nuclear reactors. %
Although originally formulated phenomenologically, the RTE can be derived from Maxwell's equations under the assumption that the wave is uncorrelated with the medium\cite{IshimaruA1978-book, MishchenkoM2014-book}.

\par
\begin{figure}[ht]%
\includegraphics{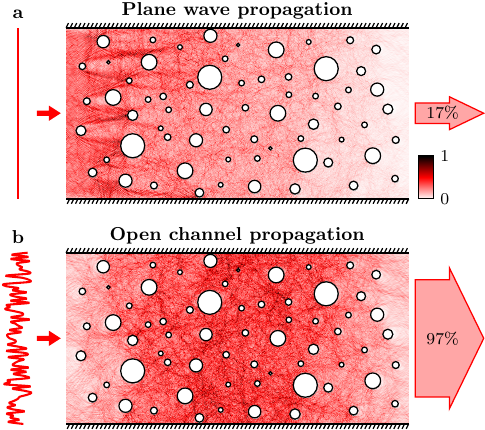}%
\caption{{\bf Plane wave versus open channel propagation.}
Intensity of a monochromatic wave propagating in a two-dimensional waveguide perturbed by random fixed obstacles.
The edges and obstacles are perfectly reflective and the space between them is scattering-free.
{\bf a}, Intensity averaged over the lowest $10$ waveguide modes.
{\bf b}, Intensity averaged over the highest $10$ transmission eigenchannels. 
In both panels, the intensity is obtained by numerically solving the wave equation (see \hyperref[sec:methods]{Methods}).
The waveguide has width $W=83.3\,\lambda$ and length $L=166.6 \,\lambda$, with mean free path $\ell=21.7\,\lambda$; it supports $166$ propagating modes.}%
\label{fig:guide-bubbles-v1}%
\end{figure}

 An important corollary of the RTE is Ohm's law, which states that the power transmitted through a scattering medium scales as $L^{-1}$, where $L$ is its thickness.
If a plane wave is sent into a waveguide perturbed by random obstacles, as shown in Fig.\ \ref{fig:guide-bubbles-v1}a, the RTE predicts a linear decrease of intensity with depth, resulting in the $L^{-1}$ scaling of transmitted power.
This law, first observed in electricity, was long believed to be universal in systems where propagation is impeded by random obstacles.

\par Experiments on optical wavefront control, however, have demonstrated that this limitation can be overcome \cite{CaoH2022}.
By shaping the incident wavefront,  it is possible to compensate for the effect of disorder in order to focus light behind an opaque material \cite{Vellekoop2007} or to maximize the power transmitted through it \cite{Vellekoop2008, KimM2012, Popoff2014, Hsu2017, Pelc2025}, as in Fig.\ \ref{fig:guide-bubbles-v1}b.
These shaped waves, known as open channels, have been extensively studied experimentally, revealing surprising properties such as enhanced energy accumulation inside the material \cite{Sarma2016, HongP2018, ShiZ2018}, transverse localization \cite{Yilmaz2019a}, a large angular memory effect \cite{Yilmaz2019b}, low fluctuations \cite{Bender2020, HongP2025}, and distinctive velocities \cite{Genack2024}.

\par On the other hand, very little is known theoretically about the propagation of shaped waves in scattering media.
Their transmission can be described using predictions from random matrix theory, originally developed in condensed matter physics \cite{Beenakker1997, Rotter2017a} and further extended to account for open systems \cite{Goetschy2013, Popoff2014, Hsu2017, Bender2022b}.
It is for instance known that nearly $100\%$ of the incident power can always be transmitted through a lossless medium, regardless of its opacity.
Apart from this, previous studies addressing propagation inside the medium \cite{Davy2015b, Ojambati2016, Koirala2017, FangP2019, vanTiggelenB2025} have relied on phenomenological approaches which, although sometimes predictive in simple structures, do not provide a general transport theory for shaped waves.
Because the behavior of open channels is so different from RTE predictions, it has long been thought that such waves escape the framework of diffusion and Brownian motion.
A long-standing question is therefore whether a transport theory can capture their unusual propagation.

\par A recently developed microscopic theory introduced a transport equation for a matrix-valued quantity carrying phase information \cite{GaspardD2025a, GaspardD2025b}.
This theory was shown to capture the distribution of transmission eigenvalues $T$, which are the eigenvalues of $\herm{\matr{t}}\matr{t}$, where $\matr{t}$ is the transmission matrix relating wavefronts from one end of a medium to the other.
The central equation is similar in structure to the RTE, but unlike the latter, the transported object is a complex-valued $2\times2$ matrix carrying wave-phase information, and the equation is nonlinear.
This nonlinearity is reminiscent of the collision term of the nonlinear Boltzmann equation, despite the underlying wave dynamics being linear.

\par Here we show that this matrix transport equation (MTE) has a much broader significance than previously recognized.
Rather than being merely an equation for transmission eigenvalues, it governs coherent wavefront propagation inside random media, providing a framework for the systematic study of shaped-wave transport.

\par To illustrate the predictive power of the MTE, we derive analytic expressions for the position-dependent energy and current densities associated with transmission eigenchannels, i.e., shaped waves achieving a given transmission $T$ inside lossless diffusive media of arbitrary shape, in excellent agreement with ab initio simulations of the wave equation.
Remarkably, these quantities can be expressed explicitly in terms of the RTE solution for unshaped waves propagating through the same domain.
We further reveal the key role of the nonlinear nature of the MTE by developing a geometric interpretation of its dynamics on a spherical manifold, providing a qualitative understanding of shaped-wave propagation and naturally extending to media with absorption and scattering loss.
Finally, we show that complex disordered structures with unusual propagation properties can be inferred from the MTE without explicitly solving the underlying wave problem.
We demonstrate this by efficiently steering waves along selected scattering paths through a simple choice of the appropriate transmission eigenchannel.

\section*{Matrix transport theory}\label{sec:theory}%
Under the assumption of a lack of correlation between a monochromatic wave (of wavelength $\lambda$) and the scattering medium, which applies in the absence of wavefront shaping (Fig.\ \ref{fig:guide-bubbles-v1}a), the radiance $f(\vect{u},\vect{r})$ obeys the RTE,
\begin{equation}\label{eq:radiative-transport}
\vect{u}\cdot\grad_{\vect{r}}f = \oint \frac{f(\vect{u}',\vect{r}) - f(\vect{u},\vect{r})}{S_d\ell} \D{\vect{u}'} - \frac{f(\vect{u},\vect{r})}{\labso} ,
\end{equation}
where $\vect{u}$ is the direction of propagation (a unit vector), $\vect{r}$ the position, $\ell$ the mean free path and $\labso$ the ballistic absorption length (both depend on $\lambda$), and $S_d=\oint\D{\vect{u}}$ a normalization factor that depends on the spatial dimension $d$.
The RTE states that the local variation of radiance in the direction $\vect{u}$ results from the imbalance between scattering into and out of $\vect{u}$, as well as from local absorption.
Note that Eq.\ \eqref{eq:radiative-transport} assumes isotropic scattering, but can be generalized to direction-dependent scattering by replacing $1/S_d$ with the appropriate phase function.

\par When the incident wavefront is shaped to maximize transmission, the phase relations between scattering paths are no longer washed out, and propagation falls outside the scope of the RTE \eqref{eq:radiative-transport} (Fig.\ \ref{fig:guide-bubbles-v1}b).
In Refs.\cite{GaspardD2025a, GaspardD2025b}, we introduced a MTE to describe this regime and showed that it captures the transmission eigenvalue distribution $\rho(T)$.
The MTE reads
\begin{equation}\label{eq:eilenberger-full}
\vect{u}\cdot\grad_{\vect{r}}\matr{g} = \oint \frac{[\matr{g}(\vect{u},\vect{r}), \matr{g}(\vect{u}',\vect{r})]}{2S_d\ell} \D{\vect{u}'} - \frac{[\matr{\sigma}_3, \matr{g}(\vect{u},\vect{r})]}{2\labso} ,
\end{equation}
where $[\matr{A}, \matr{B}] = \matr{A}\matr{B}-\matr{B}\matr{A}$ is the matrix commutator, and $\matr{\sigma}_3 = (\begin{smallmatrix}1 & 0\\ 0 & -1\end{smallmatrix})$.
This equation is structurally similar to the RTE and is still parametrized by $\ell$ and $\labso$ only, but presents key differences.
First, the radiance $\matr{g}(\vect{u},\vect{r})$ is a complex $2\times2$ matrix, whereas $f(\vect{u},\vect{r})$ is a positive scalar.
Second, the scattering imbalance is nonlinear in $\matr{g}(\vect{u},\vect{r})$, as if matrix radiances from different directions were interacting with one another.
This nonlinearity is responsible for the involutive constraint $\matr{g}(\vect{u},\vect{r})^2 = \matr{1}_2$, which implies that $\matr{g}(\vect{u},\vect{r})$ contains only four independent real variables.
This constraint strongly impacts the dynamics, as demonstrated below, precluding a perturbative treatment of the nonlinearity.
A similar transport equation has also been derived in superconductivity, where the matrix structure originates from the electron-hole degree of freedom \cite{Eilenberger1968, Usadel1970}.

\par The transmission eigenvalue distribution can be derived from the flux integral over the output surface $\mathcal{S}_\cout$ of the upper-right element of the matrix current $\vect{\matr{J}}(\vect{r})=\avg{\vect{u}\,\matr{g}(\vect{u},\vect{r})}_{\vect{u}}$, namely $\rho(T)\propto\int_{\mathcal{S}_\cout} \D{\vect{y}}\cdot\Re\vect{J}_{12}(\vect{r})$.
The dependence of $\matr{g}(\vect{u},\vect{r})$ on $T$ arises from the boundary conditions associated with the MTE \cite{GaspardD2025b}.

\par The key advance of the present work is to show that $\matr{g}(\vect{u},\vect{r})$ has a physical meaning throughout the scattering medium, at all positions and directions, rather than serving only as an intermediary for obtaining $\rho(T)$.
In particular, the average intensity profile of transmission eigenchannels is given by the upper-right element of the matrix $\matr{Q}(\vect{r})=\avg{\matr{g}(\vect{u},\vect{r})}_{\vect{u}}$ according to $I_{T}(\vect{r})\propto\Re Q_{12}(\vect{r})$.
More generally, the average radiance of transmission eigenchannels derives from $f_T(\vect{u},\vect{r})\propto\Re g_{12}(\vect{u},\vect{r})$ and their current density from $\vect{j}_{T}(\vect{r}) \propto \Re\vect{J}_{12}(\vect{r})$ (see \hyperref[sec:methods]{Methods} and Supplementary Sec.\ \ref{app:theory-radiance}). 

\par In the diffusive regime, where the propagation distance greatly exceeds $\ell$, Eq.\ \eqref{eq:eilenberger-full} reduces to a nonlinear matrix diffusion equation (see Supplementary Sec.\ \ref{app:usadel-derivation}),
\begin{equation}\label{eq:usadel-full}
\grad_{\vect{r}}\cdot\vect{\matr{J}}(\vect{r}) = -\frac{[\matr{\sigma}_3, \matr{Q}(\vect{r})]}{2\labso} , \quad
\vect{\matr{J}}(\vect{r}) = -\frac{\ell}{d} \matr{Q}(\vect{r}) \grad_{\vect{r}}\matr{Q}(\vect{r}) ,
\end{equation}
accompanied by the constraint $\matr{Q}(\vect{r})^2 = \matr{1}_2$.
The matrix current $\vect{\matr{J}}(\vect{r})$, which is a $d$-vector of matrices, obeys a nonlinear Fick's law because Eq.\ \eqref{eq:eilenberger-full} is itself nonlinear.
Equation \eqref{eq:usadel-full} must also include boundary conditions that explicitly depend on $T$ (see \hyperref[sec:methods]{Methods}).

\par Equation \eqref{eq:usadel-full} captures richer physics than the classical diffusion equation, which is limited to Ohmic transmission $\propto\frac{\ell}{L}$.
First, it predicts without further restriction the bimodal law, namely that, in a lossless medium, the transmission eigenvalue distribution is $\rho(T)=\frac{\bar{T}}{2T\sqrt{1-T}}$ for $0<T<1$, where $\bar{T}\propto\frac{\ell}{L}$ is the mean transmission.
This result holds regardless of the shape of the diffusive medium \cite{Nazarov1994a} or the number of controlled surfaces, provided that injection and output channels are distinct (see Supplementary Sec.\ \ref{app:bimodal-any-structure}).

\par
\begin{figure}[ht]%
\includegraphics{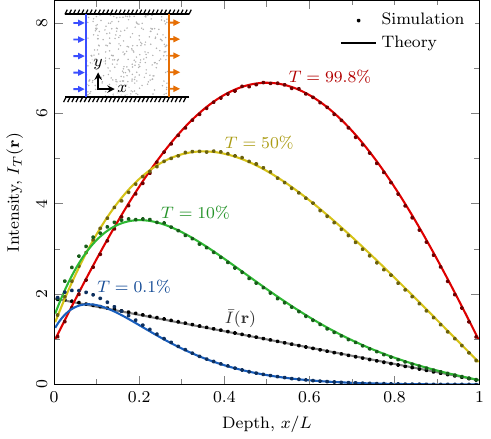}%
\caption{{\bf Transmission eigenchannel profiles in a disordered waveguide.}
Intensity profiles of transmission eigenchannels, averaged over the cross section, as a function of the longitudinal coordinate $x$.
Dots show numerical simulations based on the wave equation, averaged over $1000$ realizations of the disorder, while continuous lines show the predictions of Eq.\ \eqref{eq:itprofile-sine-law}.
Profiles are shown for transmission eigenvalues $T=99.8\%$, $T=50\%$, $T=10\%$, and $T=0.1\%$.
The classical diffusive intensity $\bar{I}(\vect{r})$, corresponding to a random incident wavefront, is also shown for comparison.
The incident intensity is normalized to 1 for all curves.
The waveguide has size $L=W=50\,\lambda$ and optical thickness $L/\ell=15$, with mean transmission $\bar{T}=9.48\%$.
The inset depicts the waveguide, with incoming arrows indicating the injection region and outgoing arrows the output region.}%
\label{fig:waveguide-v3}%
\end{figure}

 Second, Eq.\ \eqref{eq:usadel-full} yields the analytical expression for the intensity $I_{T}(\vect{r})$ deposited inside the material by a transmission eigenchannel with arbitrary transmission $T$ (see Supplementary Sec.\ \ref{app:itprofile-solution}):
\begin{equation}\label{eq:itprofile-sine-law}
I_T(\vect{r}) = \frac{2T}{\pi\bar{T}} \cosh\left[ \arccosh\left( \frac{1}{\sqrt{T}} \right) \bar{I}(\vect{r}) \right] \sin\left[ \frac{\pi}{2} \bar{I}(\vect{r}) \right] .
\end{equation}
This general result is valid in arbitrary lossless diffusive media of any shape and any $d$.
It reveals an unexpected relationship between $I_{T}(\vect{r})$ and $\bar{I}(\vect{r})$, the intensity resulting from a random wavefront launched from the same injection channels, both with their average incident intensity normalized to one (see \hyperref[sec:methods]{Methods}).
The latter satisfies the classical diffusion equation $\lapl_{\vect{r}}\bar{I}(\vect{r})=0$, with the boundary conditions $\bar{I}_\cin=2$ at the input and $\bar{I}_\cout=0$ at the output.
In Fig.\ \ref{fig:waveguide-v3}, the prediction \eqref{eq:itprofile-sine-law} is validated for a straight two-dimensional disordered waveguide across various values of $T$ against finite-element simulations of the wave equation (see \hyperref[sec:methods]{Methods}).
The $\cosh$ function in Eq.\ \eqref{eq:itprofile-sine-law} is responsible for an attenuation of the intensity for $T<1$.
At perfect transmission, $T=1$, this prefactor disappears, and the profile reduces to a pure sine wave:
$I_1(\vect{r}) = \frac{2}{\pi\bar{T}} \sin\left[ \frac{\pi}{2} \bar{I}(\vect{r}) \right]$.
Since the cross section of the waveguide does not vary in Fig.\ \ref{fig:waveguide-v3}, the random-input intensity is a linearly decreasing ramp, $\bar{I}(\vect{r}) = 2\frac{L+\ell-x}{L+2\ell}$, where $x$ is the depth.
This results in a bell-shaped profile for $I_1(\vect{r})$ centered in the middle of the waveguide, where $\bar{I}(\vect{r})\simeq 1$. 

\par A previous phenomenological approach to the profile of the open channel $I_1(\vect{r})$ interpreted it as the return probability to position $\vect{r}$ for a classical diffusive process \cite{Davy2015b}.
Although this picture compares reasonably well with simulations in various structures \cite{Koirala2017, ShiZ2018, FangP2019}, it predicts a parabolic profile rather than the sinusoidal profile from Eq.\ \eqref{eq:itprofile-sine-law}.

\par It is also worth noting that \eqref{eq:itprofile-sine-law} contains no adjustable parameter, with the exception of $\ell$.
The discrepancies observed near the input at low transmission ($T\ll\bar{T}$) in Fig.\ \ref{fig:waveguide-v3} are attributed to the diffusion approximation made between Eqs.\ \eqref{eq:eilenberger-full} and \eqref{eq:usadel-full}.

\section*{Geometric interpretation: Transport on a sphere}\label{sec:sphere}%
\par
\begin{figure}[ht]%
\includegraphics{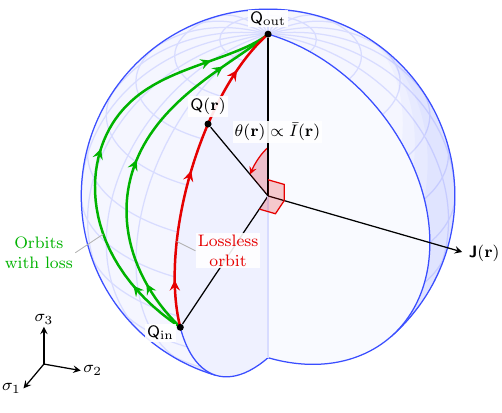}%
\caption{{\bf Spherical manifold of $\matr{Q}(\vect{r})$.}
Schematic representation of the manifold of $\matr{Q}(\vect{r})$ resulting from the involutive constraint $\matr{Q}(\vect{r})^2=\matr{1}_2$ when projected onto the three Pauli matrices $\matr{\sigma}_1,\matr{\sigma}_2,\matr{\sigma}_3$.
In a lossless waveguide, Eq.\ \eqref{eq:usadel-full} admits a solution for $\matr{Q}(\vect{r})$ confined to a great circle connecting $\matr{Q}_{\cin}$ and $\matr{Q}_{\cout}$ (red curve), orthogonal to all the spatial components of the matrix current $\vect{\matr{J}}(\vect{r})$, the latter being collinear in matrix space.
In the presence of loss, the trajectory of $\matr{Q}(\vect{r})$ depends on the path followed (green curves) but remains constrained to the sphere (see Supplementary Sec.\ \ref{app:spherical-transport}).
In this case, the spatial components of $\vect{\matr{J}}(\vect{r})$ are not collinear in general.}%
\label{fig:qn-manifold-simple-v1}%
\end{figure}

 What, then, is the origin of the sinusoidal profile of open channels predicted by Eq.\ \eqref{eq:itprofile-sine-law}? The answer lies in geometry.
Projecting the matrix quantities $\matr{Q}(\vect{r})$ and $\vect{\matr{J}}(\vect{r})$ onto the basis of the three Pauli matrices, as illustrated in Fig.\ \ref{fig:qn-manifold-simple-v1}, turns them into three-component vectors in matrix space.
The constraint $\matr{Q}(\vect{r})^2 = \matr{1}_2$ confines $\matr{Q}(\vect{r})$ to a unit sphere, while the condition $\vect{\matr{J}}(\vect{r})\matr{Q}(\vect{r}) + \matr{Q}(\vect{r})\vect{\matr{J}}(\vect{r}) = 0$ (see Supplementary Sec.\ \ref{app:usadel-derivation}) implies that $\matr{Q}(\vect{r})$ is orthogonal to each spatial component of the current.
In a lossless waveguide, Eq.\ \eqref{eq:usadel-full} admits a solution for which $\matr{Q}(\vect{r})$ remains confined to the great circle passing through the input and output boundary conditions $\matr{Q}_\cin$ and $\matr{Q}_\cout$.
The matrix current then has a fixed direction in matrix space, normal to this great circle, while its spatial dependence follows the classical diffusive current $\bar{\vect{j}}(\vect{r})$.
The motion of $\matr{Q}(\vect{r})$ along this great circle is therefore described by a single angular coordinate, $\theta(\vect{r})$, directly proportional to the classical diffusive intensity, $\bar{I}(\vect{r})$.
Taking the real part of the upper-right matrix element of $\vect{\matr{J}}(\vect{r})$ yields $\vect{j}_{T}(\vect{r})=(T/\bar{T})\bar{\vect{j}}(\vect{r})$ (see Supplementary Sec.\ \ref{app:jtcurrent-solution}), while doing the same for $\matr{Q}(\vect{r})$ gives Eq.\ \eqref{eq:itprofile-sine-law}.
Therefore, the sinusoidal profile of open channels ultimately stems from the spherical geometry imposed by the involutive constraint.

\par This spherical manifold should not be confused with the Bloch sphere of a spin-$\frac{1}{2}$ system: here, the matrix $\matr{Q}(\vect{r})$ is an auxiliary matrix describing wave transport rather than a physical two-level state.
Importantly, the constraint $\matr{Q}(\vect{r})^2 = \matr{1}_2$ is preserved even in the presence of losses or absorption.
In the latter case, the current is no longer conserved, $\matr{Q}(\vect{r})$ can deviate from the great circle (green curves in Fig.\ \ref{fig:qn-manifold-simple-v1}), and Eq.\ \eqref{eq:usadel-full} no longer admits an analytical solution in terms of elementary functions, as we shall see in the examples of the following sections.
In Supplementary Sec.\ \ref{app:spherical-transport}, we also point out an analogy with the equation of motion of the spherical pendulum, where space plays the role of time and absorption plays the role of gravity.

\par In the literature \cite{Ojambati2016, vanTiggelenB2025}, it has been suggested that the sinusoidal profile originates from the first eigenmode of the classical diffusion equation.
Here, however, we demonstrate that it has a nonlinear origin (see Supplementary Sec.\ \ref{app:irreducibility-to-diffusion}).

\section*{Open channels in complex structures}\label{sec:complex-structures}%

\par
\begin{figure*}[ht]%
\includegraphics{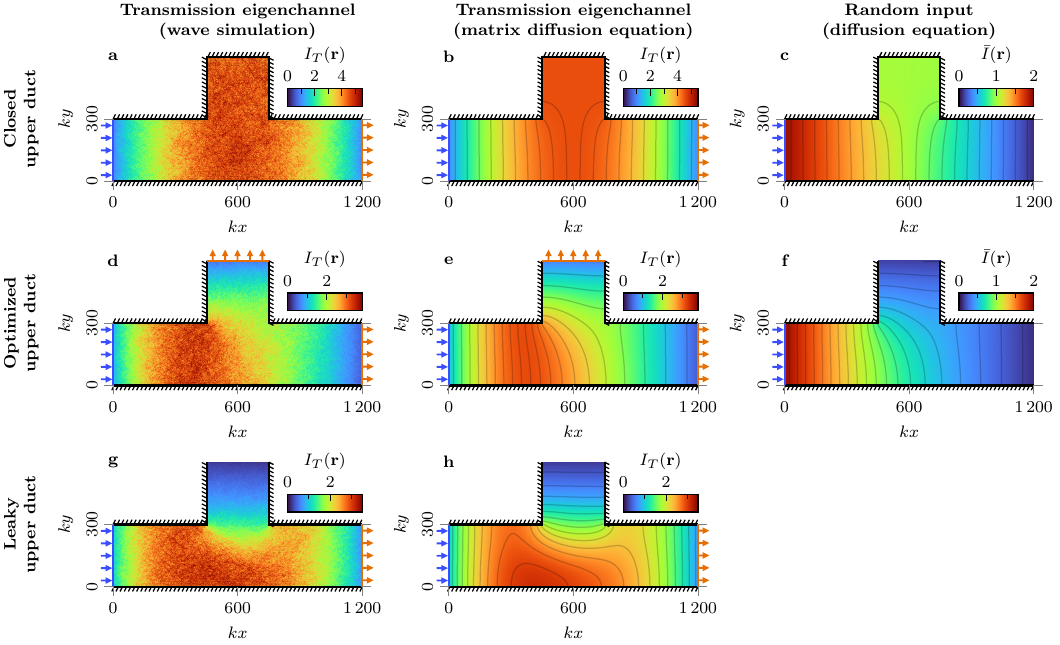}%
\caption{{\bf Influence of boundary conditions on open channels.}
Wavefront control in a two-dimensional disordered waveguide with optical thickness $L/\ell=10$ in the horizontal direction.
Incoming arrows indicate the controlled input region, while outgoing arrows indicate the target region where the transmission is maximized.
{\bf a}, {\bf b}, Open channel  with $T=1$ when the upper duct is closed and perfectly reflecting.
{\bf d}, {\bf e}, Open channel  with $T=1$ when transmission is maximized through both the upper and right ducts.
{\bf g}, {\bf h}, Open channel when the upper duct is leaky and excluded from transmission optimization, resulting in a maximum transmission of $T=70\%$.
Numerical simulations based on the wave equation, averaged over $100$ realizations, are shown in {\bf a}, {\bf d}, {\bf g}; predictions of Eq.\ \eqref{eq:itprofile-sine-law} are shown in {\bf b}, {\bf e} for $T=1$, and the numerical solution of Eq.\ \eqref{eq:usadel-full} in {\bf h} for $T=70\%$.
{\bf c}, {\bf f}, Theoretical intensity $\bar{I}(\vect{r})$ for a random, unshaped wavefront, satisfying the diffusion equation $\lapl_{\vect{r}}\bar{I}(\vect{r})=0$.}%
\label{fig:branched-guide-v1}%
\end{figure*}

 The geometric interpretation above has a direct physical consequence: the spatial profile of an open channel is controlled not only by the disorder, but also by the shape and boundary conditions of the scattering region.
We illustrate this principle in a branched disordered waveguide, where changing the boundary conditions alone produces qualitatively different open-channel profiles (Fig.\ \ref{fig:branched-guide-v1}).

\par When the upper duct is closed, the classical intensity $\bar{I}(\vect{r})$ develops a plateau halfway through the structure (Fig.\ \ref{fig:branched-guide-v1}c), which originates from the zero-flux boundary condition at walls, $\vect{n}\cdot\grad_{\vect{r}}\bar{I}(\vect{r})=0$.
For $T=1$, Eq.\ \eqref{eq:itprofile-sine-law} maps this plateau onto the open-channel intensity, as demonstrated by the numerical simulation in Fig.\ \ref{fig:branched-guide-v1}a and the theoretical prediction in Fig.\ \ref{fig:branched-guide-v1}b.
More generally, because $I_{T}(\vect{r})$ is a function of $\bar{I}(\vect{r})$, the two intensities have identical level lines, even though their values differ.

\par When the upper duct is opened and transmission through both the upper and right ducts are simultaneously maximized, the peak shifts toward the input (Fig.\ \ref{fig:branched-guide-v1}d).
This behavior is again predicted by Eq.\ \eqref{eq:itprofile-sine-law} for $T=1$ (Fig.\ \ref{fig:branched-guide-v1}e).
In fact, the peak of $I_{1}(\vect{r})$ always occurs where $\bar{I}(\vect{r})$ reaches half its maximum value (Fig.\ \ref{fig:branched-guide-v1}f).

\par The theory also describes situations in which this mapping between $I_{T}(\vect{r})$ and $\bar{I}(\vect{r})$ breaks down.
When the upper duct is open but excluded from transmission optimization, the open channel tends to avoid it (Fig.\ \ref{fig:branched-guide-v1}g).
Equation \eqref{eq:usadel-full} must then be solved numerically, yielding the prediction in Fig.\ \ref{fig:branched-guide-v1}h.
The level lines of $I_{T}(\vect{r})$ differ from those of $\bar{I}(\vect{r})$ (Fig.\ \ref{fig:branched-guide-v1}f), demonstrating that the MTE contains information beyond the classical diffusive intensity landscape.

\par
\begin{figure}[ht]%
\includegraphics{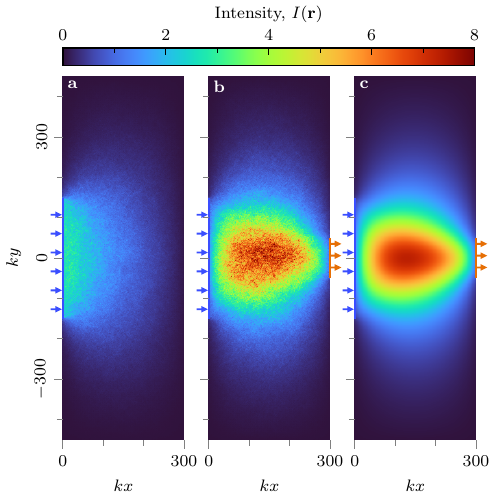}%
\caption{{\bf Wave focusing through a disordered slab.}
Wave focusing through a two-dimensional disordered slab of optical thickness $L/\ell=20$, with the wave free to escape from all four edges.
{\bf a}, Intensity for an unshaped collimated beam. Transmission to the target region (orange) is $T=1.7\%$.
{\bf b}, Intensity profile of the highest transmission eigenchannel. Transmission to the target region is $T=35\%$.
Numerical simulations based on the wave equation, averaged over $100$ realizations, are shown in {\bf a} and {\bf b}; prediction of Eq.\ \eqref{eq:usadel-full} for $T=35\%$ is shown in {\bf c}.}%
\label{fig:slab-transmission-v12}%
\end{figure}

 This predictive capability extends beyond waveguides to open disordered slabs, where waves can escape at the boundaries (Fig.\ \ref{fig:slab-transmission-v12}).
Such systems are standard in optical wavefront-shaping experiments \cite{Vellekoop2007, Vellekoop2008, KimM2012, Popoff2014, Hsu2017} .
When the blue region is illuminated with an unshaped collimated beam, the fraction of power transmitted into the orange region is only $T=1.7\%$ (Fig.\ \ref{fig:slab-transmission-v12}a).
Optimizing the incident wavefront increases this transmission to $T=35\%$ (Fig.\ \ref{fig:slab-transmission-v12}b).
The profile predicted by Eq.\ \eqref{eq:usadel-full} agrees with the numerical simulation (Fig.\ \ref{fig:slab-transmission-v12}b, c).
The theory  predicts both the longitudinal and transverse profiles of the transmission eigenchannels, as well as their eigenvalue distribution, for different sizes of the illuminated and target regions, from localized energy focusing to energy enhancement over extended target regions (see Supplementary Sec.\ \ref{app:slab-detail}).

\section*{Symmetry breaking of open channels}\label{sec:symmetry-breaking}%

\par Matrix transport theory further reveals how open channels respond to different sources of symmetry breaking.
While structural asymmetry does not strongly affect how highly transmitting eigenchannels propagate, a local perturbation that breaks the symmetry through absorption can have a dramatic effect, owing to the coherent nature of eigenchannels.

\par
\begin{figure*}[ht]%
\includegraphics{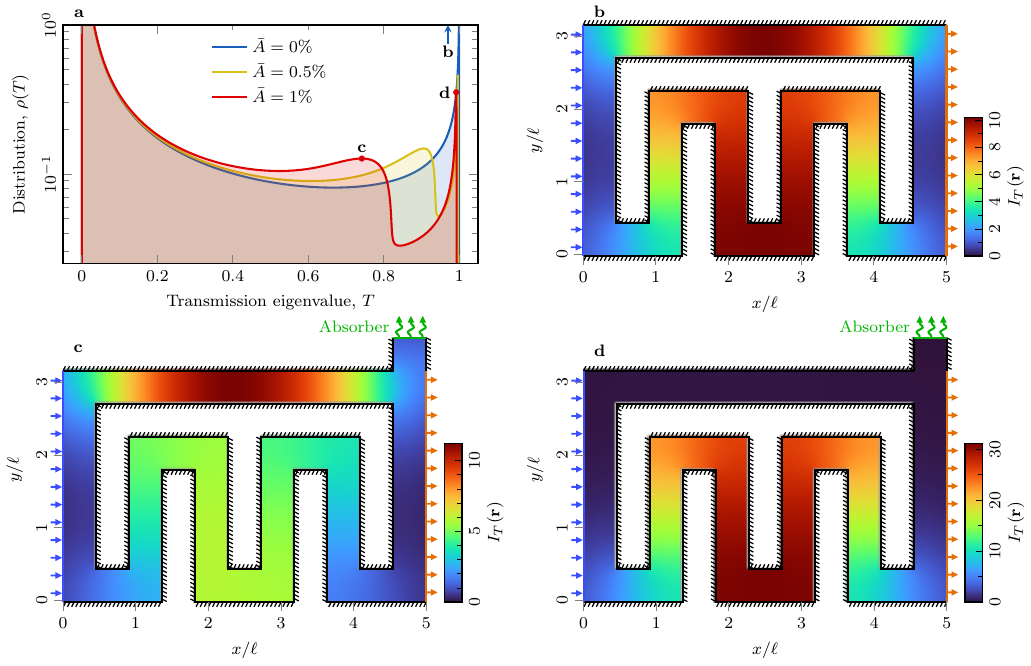}%
\caption{{\bf Symmetry breaking of open channels induced by absorption.}
{\bf a}, Distribution of transmission eigenvalues predicted by Eq.\ \eqref{eq:usadel-full} in a two-duct waveguide with spatially uniform disorder strength.
The upper duct is short ($L/\ell=5$) but potentially leaky because of an absorber, while the lower duct is long but lossless.
A pseudo-gap in transmission opens as $\bar{A}$, the mean power fraction dissipated by the absorber (see \hyperref[sec:methods]{Methods}), increases.
{\bf b}, Transmission eigenchannel intensity for $T=1$ without absorption ($\bar{A}=0\%$), predicted by Eq.\ \eqref{eq:itprofile-sine-law}.
{\bf c}, Transmission eigenchannel intensity for $T=74\%$ with $\bar{A}=1\%$ absorption, predicted by Eq.\ \eqref{eq:usadel-full}.
{\bf d}, Transmission eigenchannel intensity for $T=99\%$ with $\bar{A}=1\%$ absorption, predicted by Eq.\ \eqref{eq:usadel-full}.}%
\label{fig:circuit-v2}%
\end{figure*}

 We illustrate this effect in a two-duct disordered circuit, as shown in Fig.\ \ref{fig:circuit-v2}b.
The lower duct is more than two times longer than the upper one.
In the absence of absorption, however, the open channel is not preferentially concentrated in the shorter duct.
Despite the asymmetric structure, Eq.\ \eqref{eq:itprofile-sine-law} predicts the same peak intensity in both ducts, independently of $T$,
as shown for $T=1$ in Fig.\ \ref{fig:circuit-v2}b.
Correspondingly, $\rho(T)$ remains bimodal and is parametrized only by the $\bar{T}$ of the whole structure (blue line in Fig.\ \ref{fig:circuit-v2}a).

\par A qualitatively different behavior emerges when a small absorber is introduced near the exit of the short duct.
The eigenvalue distribution predicted by Eq.\ \eqref{eq:usadel-full} develops a trimodal shape with a pronounced pseudo-gap at high transmission (yellow and red lines in Fig.\ \ref{fig:circuit-v2}a).
Below the gap, an eigenchannel with $T=74\%$ preferentially propagates through the short duct (Fig.\ \ref{fig:circuit-v2}c), whereas above the gap, an eigenchannel with $T=99\%$ preferentially propagates through the long duct (Fig.\ \ref{fig:circuit-v2}d).
Numerical simulations of the wave equation confirm this result (see Supplementary Sec.\ \ref{app:circuit-check}).

\par This behavior shows that highly transmitting eigenchannels with similar transmission can preferentially couple to very different paths.
By selecting the appropriate eigenchannel, one can therefore steer waves through a complex disordered system by exploiting symmetry breaking, without inverse-designing the microscopic disorder or solving the full wave equation.

\section*{Discussion}\label{sec:discussion}%

\par We have established a transport theory for shaped waves in disordered media, in which the transported quantity is an involutive matrix rather than a scalar radiance.
In lossless media, the nonlinear MTE yields analytical predictions for transmission eigenchannels, including the sinusoidal intensity profile of open channels and $\cosh$-shaped evanescence of closed channels, and reveals their geometric origin.
It also clarifies the relation to previous interpretations of the open-channel profile based on a return probability \cite{Davy2015b, Koirala2017, FangP2019} or on the first eigenmode of the diffusion equation \cite{Ojambati2016, vanTiggelenB2025}: the former leads to a parabolic profile, while the latter does not capture the nonlinear origin identified here.
Beyond simple waveguides, the same framework describes complex structures, boundary conditions, and absorption, and reveals how symmetry breaking can reorganize highly transmitting eigenchannels.

\par The MTE provides an efficient framework for studying wavefront control in structures much larger than the wavelength, without wavelength-scale discretization while retaining the information required for coherent optimization.
This opens the possibility of exploring transmission eigenchannels in complex optical circuits containing random inclusions or defects, and of designing structures with prescribed transmission properties \cite{Molesky2018}.
The framework may also be extended to inhomogeneous or anisotropically scattering media, such as biological tissues \cite{Horstmeyer2015, ParkJ2018}.

\par The matrix formulation gives access to observables beyond bulk intensity, including current, energy velocity \cite{Genack2024, Joshi2024}, and radiance, which are particularly relevant in anisotropic media.
An important extension would be to develop two-point intensity correlations, which could capture phenomena washed out by disorder averaging, such as transverse localization of transmission eigenchannels \cite{Yilmaz2019a}.
The framework could likewise be extended to other optimization objectives, such as energy deposition \cite{ChengX2014, Horstmeyer2015, Bender2022a}, and to time-dependent wavefront control of pulses \cite{ChoiY2013, ShiZ2015b, JeongS2018, Devaud2022}.

\par Several fundamental questions remain open.
One is whether the MTE can be extended to the Anderson localized regime using a self-consistent renormalization approach \cite{WolfleP2010}.
Another concerns the intrinsic nonlinearity of the MTE: whether the constrained evolution of $\matr{Q}$ on the sphere can support nontrivial structures such as solitons, and whether some may admit a topological characterization.
These questions could reveal new manifestations of nonlinear dynamics generated by an underlying linear wave equation.

\par More broadly, our results show that coarse-grained transport can retain information about deliberately controlled coherence, extending transport theory to regimes where wavefront control plays an essential role, be it for light, sound, or any type of wave.

\bibliography{article-profile-theory}%

\clearpage%
\section*{Methods}\label{sec:methods}%

\subsection*{Wave propagation simulations}%

\par Here, we present the method used in the wave simulations of Figs.\ \ref{fig:guide-bubbles-v1}, \ref{fig:waveguide-v3}, \ref{fig:branched-guide-v1}, and \ref{fig:slab-transmission-v12}.
These simulations are performed by the Waffle program\cite{GaspardD2026-waffle} using the finite-element method.
The starting point of all wave simulations is the computation of the retarded modal Green's function $G^+_{n}(\vect{r})$ defined by
\begin{equation}\label{eq:green-input-modes}
\left( \lapl_{\vect{r}} + k^2 + \I\varepsilon - U(\vect{r}) \right) G^+_{n}(\vect{r}) = \chi_{\cin,n}(\vect{y}) \delta(x-x_\cin) ,
\end{equation}
where $k$ is the wavenumber, $\varepsilon=k/\labso$ the absorption parameter, and $U(\vect{r})$ represents the disordered potential. 

\par In Fig.\ \ref{fig:guide-bubbles-v1}, $U(\vect{r})$ describes perfectly reflecting obstacles, so that it is zero between them and infinitely large inside, while in Figs.\ \ref{fig:waveguide-v3}, \ref{fig:branched-guide-v1}, and \ref{fig:slab-transmission-v12}, $U(\vect{r})$ is a delta-correlated Gaussian noise with $\tavg{U(\vect{r})U(\vect{r}')}=\alpha\delta(\vect{r}-\vect{r}')$, whose magnitude $\alpha$ determines the value of the mean free path, $\ell=\frac{k}{\pi\nu\alpha}$, where $\nu=\frac{S_d k^{d-2}}{2(2\pi)^d}$ is the density of states per unit $k^2$.

\par The right-hand side of Eq.\ \eqref{eq:green-input-modes} is the source term located at the longitudinal coordinate $x_\cin$ in the input duct.
In this term, $\chi_{\cin,n}(\vect{y})$ is the $n$-th transverse eigenmode of the input duct defined by $(\lapl_{\vect{y}} + k_{\perp,n}^2)\chi_{\cin,n}(\vect{y}) = 0$ and normalized by $\int_{\mathcal{S}_\cin} \D{\vect{y}}\, \cc{\chi}_{\cin,m}(\vect{y}) \chi_{\cin,n}(\vect{y}) = \delta_{mn}$, where $\mathcal{S}_\cin$ is the input surface.
In the simulations, we use Dirichlet boundary conditions at the transverse edges of the input duct, so that the modes $\chi_{\cin,n}(\vect{y})$ are sine waves.
Similar definitions apply to the output eigenmodes $\chi_{\cout,n}(\vect{y})$.
At the end of each duct, we use a free-escape boundary condition ensuring that the wave leaves the system without reflection, $ \vect{n}\cdot\grad_{\vect{r}} G^+_n(\vect{r}) = \I\sqrt{k^2 + \lapl_{\vect{y}}}G^+_n(\vect{r}) $, where $\vect{n}$ is the outward normal to the duct's open edge, and $\lapl_{\vect{y}}$ is the transverse Laplacian.
The square root makes the operator nonlocal along the open edge.

\par In the program\cite{GaspardD2026-waffle}, Eq.\ \eqref{eq:green-input-modes}  is discretized over a two-dimensional square lattice of step $h$.
In particular, a five-point stencil is used to approximate the Laplacian, which affects the free-space dispersion relation: $4\sin^2(\frac{1}{2}k_xh) + 4\sin^2(\frac{1}{2}k_yh) = (kh)^2$.
However, by choosing $kh\leq 1$ in our simulations, this relation deviates by less than $2.5\%$ from the continuum dispersion relation, $k_x^2+k_y^2=k^2$.

\par From the Green's function \eqref{eq:green-input-modes}, we can evaluate the field at any point within the medium resulting from the propagation of an arbitrary incident wavefront
\begin{equation}\label{eq:incident-wavefront}
\psi_\cin(\vect{y}) = \mathcal{N} \sum_{n=1}^{N_\cin} a_n \chi_{\cin,n}(\vect{y}) ,\quad
\mathcal{N} = \left( \sum_{n=1}^{N_\cin} \frac{\abs{a_n}^2}{S_\cin} \right)^{-\frac{1}{2}} .
\end{equation}
The normalization factor $\mathcal{N}$ in Eq.\ \eqref{eq:incident-wavefront} is chosen such that the average incident intensity is normalized to one: $\int_{\mathcal{S}_\cin}\D{\vect{y}} \abs{\psi_\cin(\vect{y})}^2 = S_\cin$.
The field resulting from the propagation of $\psi_\cin(\vect{y})$ then reads
\begin{equation}\label{eq:field-arbitrary}
\psi(\vect{r}) = 2\I\mathcal{N} \sum_{n=1}^{N_\cin} k_{\cin,n} a_n G^+_n(\vect{r}),
\end{equation}
where $k_{\cin,n}$ is the input longitudinal wavenumber associated with the $n$-th transverse eigenmode.
Uniform illumination across the input duct at normal incidence can be achieved by choosing $a_n=(1-(-1)^n)/n$.
Such illumination is used in Fig.\ \ref{fig:slab-transmission-v12}a.

\par From the Green's function \eqref{eq:green-input-modes} evaluated on the output surface of the output duct, we can also compute the transmission matrix $\matr{t}$.
According to the Fisher and Lee relation\cite{FisherD1981}, its elements are
\begin{equation}\label{eq:fisher-lee-mt}
t_{mn} = 2\I\sqrt{k_{\cout,m} k_{\cin,n}} \int_{\mathcal{S}_\cout} \D{\vect{y}}\, \cc{\chi}_{\cout,m}(\vect{y}) G^+_{n}(x_\cout,\vect{y}) ,
\end{equation}
where $x_\cout$ is the longitudinal coordinate of the output duct and $\mathcal{S}_\cout$ its surface.
This expression assumes no overlap between the input and output modes.

\par The central quantities of this work, namely the transmission eigenvalues and the transmission eigenchannels, are computed numerically from the singular value decomposition (SVD) of the transmission matrix, $\matr{t} = \sum_{e=1}^{N_{\rm e}} \sqrt{T_e} \vect{u}_e \herm{\vect{v}}_e$, where $0\leq T_e\leq 1$ are the so-called transmission eigenvalues, and $\vect{u}_e$ and $\vect{v}_e$ the associated singular vectors.
The column vectors $\vect{u}_e$ (of length $N_\cout$) and $\vect{v}_e$ (of length $N_\cin$) are normalized according to $\herm{\vect{u}}_{e}\vect{u}_{e'}=\delta_{ee'}$ and $\herm{\vect{v}}_{e}\vect{v}_{e'}=\delta_{ee'}$.
The transmission eigenvalue distribution $\rho(T)$ is then formally defined by
\begin{equation}\label{eq:def-tspectrum}
\rho(T) = \frac{1}{N_{\rm e}} \sum_{e=1}^{N_{\rm e}} \avg{\delta(T-T_e)} ,
\end{equation}
where $\avg{\cdots}$ denotes the average over realizations of the disorder, $N_{\rm e}=\min(N_\cin,N_\cout)$ is the rank of $\matr{t}$, $N_\cin$ and $N_\cout$ are the number of propagating modes in the input and output ducts, respectively.

\par In addition, the transmission eigenchannel $\psi_{T_e}(\vect{r})$ associated with the transmission eigenvalue $T_e$ is given by Eq.\ \eqref{eq:field-arbitrary} for $a_n = v_{ne}/\sqrt{k_{\cin,n}}$.
The average intensity profile of transmission eigenchannels is then defined by
\begin{equation}\label{eq:def-itprofile-mt}
I_{T}(\vect{r}) = \frac{\avg{ \sum_{e=1}^{N_{\rm e}} \abs{\psi_{T_e}(\vect{r})}^2 \delta(T-T_e) }}{N_{\rm e}\rho(T)} .
\end{equation}
This equation is used numerically to generate the intensity maps in Figs.\ \ref{fig:guide-bubbles-v1}, \ref{fig:waveguide-v3}, \ref{fig:branched-guide-v1}, and \ref{fig:slab-transmission-v12}.
Similarly, the average current density of transmission eigenchannels is defined by
\begin{equation}\label{eq:def-jtcurrent-mt}
\vect{j}_{T}(\vect{r}) = \frac{\avg{ \sum_{e=1}^{N_{\rm e}} \Im[ \cc{\psi}_{T_e}(\vect{r}) \grad_{\vect{r}}\psi_{T_e}(\vect{r}) ] \delta(T-T_e) }}{kN_{\rm e}\rho(T)} .
\end{equation}

\subsection*{MTE and transmission eigenchannel observables}%

\par The matrix radiance $\matr{g}(\vect{u},\vect{r})$ appearing in Eq.\ \eqref{eq:eilenberger-full} describes the propagation of a transmission eigenchannel with eigenvalue $T$.
It is defined as the Wigner transform of $\avg{\matr{\mathcal{G}}(\vect{r},\vect{r}')}$, where the $2\times 2$ matrix Green's function $\matr{\mathcal{G}}(\vect{r},\vect{r}')$ is defined by
\begin{equation}\label{eq:def-matrix-green-1-mt}\begin{aligned}
& \Big( \lapl_{\vect{r}} + k^2 + \I\varepsilon\matr{\sigma}_3 - U(\vect{r})  \\
& + \op{K}_\cin \frac{\matr{\sigma}_+}{\sqrt{T}} + \op{K}_\cout \frac{\matr{\sigma}_-}{\sqrt{T}} \Big) \matr{\mathcal{G}}(\vect{r},\vect{r}') = \matr{1}_2\delta(\vect{r}-\vect{r}')  \:,
\end{aligned}\end{equation}
where $\matr{\sigma}_3$ and $\matr{\sigma}_\pm$ are Pauli matrices, $\op{K}_\cin = \sum_{n=1}^{N_\cin} 2k_{\cin,n} \ket{x_\cin,\chi_{\cin,n}} \bra{x_\cin,\chi_{\cin,n}}$ is the input contact operator and $\op{K}_\cout$ the output one.
After disorder averaging, the random potential $U(\vect{r})$ is replaced by the matrix self-energy $\tfrac{k}{\ell\pi\nu} \avg{\matr{\mathcal{G}}(\vect{r},\vect{r})}$.
Taking the Wigner transform and assuming slow spatial variation of $\matr{g}(\vect{u},\vect{r})$ then leads to Eq.\ \eqref{eq:eilenberger-full} in the bulk.
The effect of the contact terms, which carry the $T$-dependence, is implicitly contained in the boundary conditions for $\matr{g}(\vect{u},\vect{r})$ (see Ref.\ \cite{GaspardD2025b}).

\par The average radiance of transmission eigenchannels, $f_{T}(\vect{u},\vect{r})$, is defined as the Wigner transform of the two-point local density of transmission eigenchannels,
\begin{equation}\label{eq:def-dote-mt}
D_T(\vect{r},\vect{r}') = \frac{\avg{ \sum_{e=1}^{N_{\rm e}} \psi_{T_e}(\vect{r}) \cc{\psi}_{T_e}(\vect{r}') \delta(T-T_e) }}{N_{\rm e}\rho(T)} .
\end{equation}
We show in Supplementary Sec.\ \ref{app:theory-radiance} that it can be expressed in terms of the upper-right element of $\matr{g}(\vect{u},\vect{r})$ as
\begin{equation}\label{eq:tradiance-from-g-mt}
f_{T}(\vect{u},\vect{r}) = \frac{S_\cin}{\pi S_{\rm e}\rho(T)\sqrt{T}} \Re g_{12}(\vect{u},\vect{r}) ,
\end{equation}
where  $S_{\rm e}=\min(S_\cin,S_\cout)$.
In particular, Eq.\ \eqref{eq:tradiance-from-g-mt} provides access not only to the intensity profile of transmission eigenchannels defined by Eq.\ \eqref{eq:def-itprofile-mt},
\begin{equation}\label{eq:itprofile-from-qn-mt}
I_{T}(\vect{r}) = \oint \frac{\D{\vect{u}}}{S_d} f_T(\vect{u},\vect{r}) = \frac{S_\cin}{\pi S_{\rm e}\rho(T)\sqrt{T}} \Re Q_{12}(\vect{r}) ,
\end{equation}
where $\matr{Q}(\vect{r}) = \oint \frac{\D{\vect{u}}}{S_d} \matr{g}(\vect{u},\vect{r})$, but also to the current density of transmission eigenchannels defined by Eq.\ \eqref{eq:def-jtcurrent-mt},
\begin{equation}\label{eq:jtcurrent-from-jn-mt}
\vect{j}_{T}(\vect{r}) = \oint \frac{\D{\vect{u}}}{S_d} \vect{u}\, f_T(\vect{u},\vect{r}) = \frac{S_\cin}{\pi S_{\rm e}\rho(T)\sqrt{T}} \Re \vect{J}_{12}(\vect{r}) ,
\end{equation}
where $\vect{\matr{J}}(\vect{r}) = \oint \frac{\D{\vect{u}}}{S_d} \vect{u}\,\matr{g}(\vect{u},\vect{r})$.

\par We also show in Supplementary Sec.\ \ref{app:theory-distribution} that the distribution of transmission eigenvalues \eqref{eq:def-tspectrum} can be expressed in terms of the matrix radiance $\matr{g}(\vect{u},\vect{r})$, according to
\begin{equation}\label{eq:tspectrum-from-jn-mt}
\rho(T) = \frac{d\mu}{\pi S_{\rm e}T^{\frac{3}{2}}} \Re \int_{\mathcal{S}_\cout} \D{\vect{y}}\cdot\vect{J}_{12}(\vect{r}) ,
\end{equation}
where $\mu=\frac{V_d}{2V_{d-1}}$ is the mean direction cosine and $V_d$ is the volume of the unit ball in $\mathbb{R}^d$.
In Eq.\ \eqref{eq:tspectrum-from-jn-mt}, the current is integrated over the output surface $\mathcal{S}_\cout$, with the surface element $\D{\vect{y}}$  oriented along the outward normal to the medium.
The resulting distribution $\rho(T)$ is normalized by $\int_{0^+}^1 \D{T}\,\rho(T)=1$, excluding transmission eigenvalues exactly equal to zero.
Equation \eqref{eq:itprofile-from-qn-mt} is used in Figs.\ \ref{fig:waveguide-v3}, \ref{fig:branched-guide-v1}, \ref{fig:slab-transmission-v12}, and \ref{fig:circuit-v2}, and Eq.\ \eqref{eq:tspectrum-from-jn-mt} is used in Fig.\ \ref{fig:circuit-v2}a.

\subsection*{Matrix diffusion equation and solutions}%

\par In the diffusive regime, the matrix radiance $\matr{g}(\vect{u},\vect{r})$ can be expanded as $\matr{g}(\vect{u},\vect{r}) \simeq \matr{Q}(\vect{r}) + d\vect{u}\cdot\vect{\matr{J}}(\vect{r})$, so that the MTE \eqref{eq:eilenberger-full} is transformed into the nonlinear matrix diffusion equation \eqref{eq:usadel-full} (see Supplementary Sec.\ \ref{app:usadel-derivation}). 

\par Different boundary conditions apply to the different surfaces of the scattering medium.
At the input and output surfaces that define the transmission matrix in Eq.\ \eqref{eq:fisher-lee-mt}, the boundary conditions for the eigenchannel of transmission $T$ read (see Supplementary Sec.\ \ref{app:usadel-boundary}):
\begin{equation}\label{eq:usadel-bdr-inout-mt}\begin{aligned}
\matr{Q}(\vect{r}_\cin  + z_0\vect{n}) & = \matr{Q}_\cin  = \begin{pmatrix}1 & \tfrac{-2\I}{\sqrt{T}}\\ 0 & -1\end{pmatrix} ,\\
\matr{Q}(\vect{r}_\cout + z_0\vect{n}) & = \matr{Q}_\cout = \begin{pmatrix}1 & 0\\ \tfrac{-2\I}{\sqrt{T}} & -1\end{pmatrix} ,
\end{aligned}\end{equation}
where $\vect{r}_\cin$ and $\vect{r}_\cout$ denote positions on the input or output edges, respectively, and $z_0=\mu\ell$ is the diffusive extrapolation length.

\par For the remaining surfaces of the scattering domain, we consider in this work two possible scenarios, one corresponding to a free surface, where the wave can leak (or be absorbed) without being detected, and the other to a perfectly reflecting wall.
For a free surface, we show that (see Supplementary Sec.\ \ref{app:usadel-boundary})
\begin{equation}\label{eq:usadel-bdr-loss-mt}
\matr{Q}(\vect{r} + z_0\vect{n}) = \matr{\sigma}_3 = \begin{pmatrix}1 & 0\\ 0 & -1\end{pmatrix} ,
\end{equation}
for all $\vect{r}$ on the leaky edge.
On the other hand, at a reflecting wall, the normal component of the matrix flux is zero, or equivalently,
\begin{equation}\label{eq:usadel-bdr-mirror-mt}
\vect{n}\cdot\grad_{\vect{r}}\matr{Q}(\vect{r}) = 0 ,
\end{equation}
where $\vect{n}$ is the outward normal to the medium. 

\par To find the analytical solution \eqref{eq:itprofile-sine-law} in the lossless situation, we first solve Eq.\ \eqref{eq:usadel-full} with the boundary conditions \eqref{eq:usadel-bdr-inout-mt}, using the parametrization (see Supplementary Sec.\ \ref{app:lossless-solution-matrix})
\begin{equation}\label{eq:qn-param-lossless-mt}
\matr{Q}(\vect{r}) = \E^{-\frac{\I}{2}\matr{K}_2\theta(\vect{r})} \matr{Q}_\cout \E^{\frac{\I}{2}\matr{K}_2\theta(\vect{r})} ,
\end{equation}
where
\begin{equation}\label{eq:def-k2-axis-mt}
\matr{K}_2 = \frac{[\matr{Q}_\cout, \matr{Q}_\cin]}{\sqrt{[\matr{Q}_\cout, \matr{Q}_\cin]^2}} = \frac{1}{\sqrt{1-T}} \begin{pmatrix}1 & -\I\sqrt{T}\\ -\I\sqrt{T} & -1\end{pmatrix},
\end{equation}
and the angle $\theta(\vect{r})$ measures the angular position of $\matr{Q}(\vect{r})$ relative to $\matr{Q}_\cout$, along the great circle represented in Fig.\ \ref{fig:qn-manifold-simple-v1}.
The intensity profile then follows from Eq.\ \eqref{eq:itprofile-from-qn-mt}.

\par On the other hand, in the general lossy or absorbing case, we solve Eq.\ \eqref{eq:usadel-full} complemented by the boundary conditions \eqref{eq:usadel-bdr-inout-mt}, \eqref{eq:usadel-bdr-loss-mt}, and \eqref{eq:usadel-bdr-mirror-mt}, numerically.
For this, we use the parametrization
\begin{equation}\label{eq:qn-sigma-param-1-mt}
\matr{Q}(\vect{r}) = \E^{-\frac{\I}{2}\matr{\sigma}_2\varphi(\vect{r})} \E^{-\frac{\I}{2}\matr{\sigma}_1\vartheta(\vect{r})} \matr{\sigma}_3 \E^{\frac{\I}{2}\matr{\sigma}_1\vartheta(\vect{r})} \E^{\frac{\I}{2}\matr{\sigma}_2\varphi(\vect{r})} ,
\end{equation}
that restricts the dynamics to the manifold $\matr{Q}(\vect{r})^2 = \matr{1}_2$ and is numerically stable.
This parametrization by two complex-valued angles $\varphi(\vect{r})$ and $\vartheta(\vect{r})$ describes an arbitrary evolution on the two-dimensional spherical manifold.
The two nonlinear coupled equations for $\varphi(\vect{r})$ and $\vartheta(\vect{r})$ as well as their boundary conditions are given in Supplementary Sec.\ \ref{app:numerical-usadel}.
Those equations are solved numerically by the Usador program\cite{GaspardD2026-usador}, using the finite-element method on a two-dimensional square lattice and the Newton-Raphson iterative algorithm.
We then use \eqref{eq:itprofile-from-qn-mt} to generate the plots in Figs.\ \ref{fig:waveguide-v3}, \ref{fig:branched-guide-v1}, \ref{fig:slab-transmission-v12}, and \ref{fig:circuit-v2}.
Equation \eqref{eq:tspectrum-from-jn-mt}, with $\vect{\matr{J}}(\vect{r}) = -\frac{\ell}{d} \matr{Q}(\vect{r}) \grad_{\vect{r}}\matr{Q}(\vect{r})$, is also used to obtain the eigenvalue distribution shown in Fig.\ \ref{fig:circuit-v2}a.

\par Finally, in Fig.\ \ref{fig:circuit-v2}, the value of the power dissipated by the absorber, $\bar{A}$, is computed from the solution of the classical diffusion equation $\lapl_{\vect{r}}\bar{I}(\vect{r}) = 0$, with the boundary conditions in the input and output regions: $\bar{I}(\vect{r}_\cin + z_0\vect{n}) = 2$ and $\bar{I}(\vect{r}_\cout + z_0\vect{n}) = 0$.
We also use the zero-flux boundary condition $\vect{n}\cdot\grad_{\vect{r}}\bar{I}(\vect{r}) = 0$ at reflecting walls.
The value of $\bar{A}$ in Fig.\ \ref{fig:circuit-v2} is then obtained from the flux integral
\begin{equation}\label{eq:tabso-from-jbar}
\bar{A} = \frac{d\mu}{S_\cin} \int_{\mathcal{S}_{\rm a}} \D{\vect{y}}\cdot\bar{\vect{j}}(\vect{r}) , \qquad \bar{\vect{j}}(\vect{r}) = -\frac{\ell}{d} \grad_{\vect{r}}\bar{I}(\vect{r}) ,
\end{equation}
where the integration surface, $\mathcal{S}_{\rm a}$, is the cross section of the absorbing duct.
See also Supplementary Sec.\ \ref{app:lossless-solution-connection} for the derivation of Eq.\ \eqref{eq:tabso-from-jbar}.
In Fig.\ \ref{fig:circuit-v2}, $d=2$ and the widths of $\mathcal{S}_{\rm a}$ are zero, $0.14\,\ell$, and $0.45\,\ell$, with the last case corresponding to Fig.\ \ref{fig:circuit-v2}c, d.

\section*{Data availability}%
All data that support the plots within this paper and other findings of this study are available from the corresponding authors upon reasonable request.

\section*{Code availability}%
All the codes used to perform the numerical simulations are publicly available.
The wave equation is solved using the Waffle program\cite{GaspardD2026-waffle}, and the matrix diffusion equation \eqref{eq:usadel-full} is solved using the Usador program\cite{GaspardD2026-usador}.
Both programs internally call the UMFPACK software\cite{DavisT2004b}.

\section*{Acknowledgments}%
This research has been supported by the ANR project MARS\_light under reference \href{https://anr.fr/Project-ANR-19-CE30-0026}{ANR-19-CE30-0026}, by the program ``Investissements d'Avenir'' launched by the French Government.
It also received support from a grant of the Simons Foundation (No.\ 1027116).

\section*{Author contributions}%
D.G.\ developed the analytical model and performed the numerical simulations.
A.G.\ initiated the project, supervised the research and contributed to the model.
Both authors contributed to the manuscript preparation.

\section*{Competing interests}%
The authors declare no competing interests.

\section*{Additional information}%
\noindent\textbf{Supplementary information} This article is accompanied with a supplementary information document.\\[1em]
\noindent\textbf{Correspondence and requests for materials} should be addressed to David Gaspard or Arthur Goetschy.

\clearpage%
\appendix%
\setcounter{figure}{0}%
\counterwithin{figure}{section}%
\renewcommand{\thefigure}{\Alph{section}\arabic{figure}}%
\renewcommand{\theHfigure}{\Alph{section}\arabic{figure}}%
\onecolumngrid%
\begin{center}%
{\bfseries\Large Supplementary Information\par}\vspace{1ex}%
{\bfseries\large\papertitle\par}\vspace{3ex}%
{David Gaspard and Arthur Goetschy\par}\vspace{2ex}%
\end{center}%
\twocolumngrid%

In this document, we provide detailed derivations of the mathematical expressions appearing in the main text. It is structured as follows.
In Sec.\ \ref{app:theory}, we develop the general theory of shaped waves in disordered media.
This includes, in Sec.\ \ref{app:theory-radiance}, the theoretical predictions for the radiance and intensity profiles of transmission eigenchannels, as well as, in Sec.\ \ref{app:theory-distribution}, the distribution of transmission eigenvalues, all expressed in terms of the solution of a matrix transport equation.
In Sec.\ \ref{app:diffusion-approx}, we perform the diffusion approximation of the matrix transport equation (MTE), which provides in Sec.\ \ref{app:usadel-derivation} a matrix diffusion equation, and in Sec.\ \ref{app:usadel-boundary} appropriate boundary conditions.
In Sec.\ \ref{app:lossless-solution}, we derive and discuss the solution of this matrix diffusion equation in a lossless disordered medium.
The solution derived in Sec.\ \ref{app:lossless-solution-matrix} is related to the solution of the classical diffusion equation for a random input wavefront in Sec.\ \ref{app:lossless-solution-connection}.
This solution provides not only a proof of the bimodal law $\rho(T)=\frac{\bar{T}}{2T\sqrt{1-T}}$ in a lossless disordered medium of arbitrary structure in Sec.\ \ref{app:bimodal-any-structure}, but also analytic expressions of the intensity profile (in Sec.\ \ref{app:itprofile-solution}) and current density (in Sec.\ \ref{app:jtcurrent-solution}) of transmission eigenchannels.
In Sec.\ \ref{app:irreducibility-to-diffusion}, we discuss the fact that the sinusoidal profile of the transmission eigenchannels does not originate from the eigenmodes of the diffusion equation.
In Sec.\ \ref{app:spherical-transport}, we introduce a spherical parametrization of the matrix field $\matr{Q}(\vect{r})$ that is suitable for describing more general trajectories in matrix space in the presence of absorption and free (uncontrolled) surfaces.
In Sec.\ \ref{app:numerical-usadel}, we introduce an alternative parametrization based on the Pauli matrices and describe the numerical method implemented in the Usador program \cite{GaspardD2026-usador} to solve the matrix diffusion equation.
Finally, additional results and plots are presented in Sec.\ \ref{app:additional-results}, in particular the study of the influence of the illumination size in the disordered slab (Sec.\ \ref{app:slab-detail}), and the validation of the gap in the transmission eigenvalue distribution in the two-duct disordered waveguide (Sec.\ \ref{app:circuit-check}).

\section{Theory of transmission eigenchannels}\label{app:theory}%

\par Transmission eigenchannels $\psi_{T_e}(\vect{r})$ are defined as eigenstates of the matrix $\herm{\matr{t}}\matr{t}$ associated to the eigenvalues $0\leq T_e\leq 1$.
The transmission matrix $\matr{t}$ is given by the Fisher-Lee relation,
\begin{equation}\label{eq:fisher-lee}
t_{mn} = 2\I\sqrt{k_{\cout, m} k_{\cin, n}} \bra{x_\cout, \chi_{\cout,m}} \op{G}^+ \ket{x_\cin, \chi_{\cin,n}},
\end{equation}
where $\op{G}^+$ is the retarded Green's operator defined by
\begin{equation}\label{eq:def-green-formal}
\op{G}^+ = \frac{1}{\op{\lapl}_{\vect{r}} + k^2 + \I\varepsilon - U(\op{\vect{r}})},
\end{equation}
and $\chi_{\cin,n}$ ($\chi_{\cout,m}$) are the transverse modes of the input (output) duct, associated with the longitudinal wavenumbers $k_{\cin, n}$ ($k_{\cout, m}$).

\par According to the SVD $\matr{t} = \sum_{e=1}^{N_{\rm e}} \sqrt{T_e} \vect{u}_e \herm{\vect{v}}_e$, the incident wavefront $\psi_{\cin,T_e}(\vect{y})$ is directly related to the components of $\vect{v}_e$ by
\begin{equation}\label{eq:tstate-incident-wavefront}
\psi_{\cin,T_e}(\vect{y}) = \mathcal{N} \sum_{n=1}^{N_\cin} v_{ne} \frac{\chi_{\cin, n}(\vect{y})}{\sqrt{k_{\cin, n}}},
\end{equation}
where $\mathcal{N}$ is a normalization factor chosen such that the average incident intensity is normalized to one, $\int_{\mathcal{S}_\cin} \frac{\D{\vect{y}}}{S_\cin} \abs{\psi_{\cin,T_e}(\vect{y})}^2 = 1$, that is
\begin{equation}\label{eq:normalization-prefactor}
\mathcal{N} = \left( \frac{1}{S_\cin} \sum_{n=1}^{N_\cin} \frac{\abs{v_{ne}}^2}{k_{\cin, n}}\right)^{-1/2} \simeq \sqrt{\frac{N_\cin}{2\pi\nu(k)}},
\end{equation}
where $\nu(k) = \frac{S_d k^{d-2}}{2(2\pi)^d}$ is the density of states per unit $k^2$. 
Propagation of the wavefront \eqref{eq:tstate-incident-wavefront} through the complex system to position $\vect{r}$ results in the field
\begin{equation}\label{eq:field-eigenchannel}
\psi_{T_e}(\vect{r}) = 2\I\mathcal{N} \sum_{n=1}^{N_\cin} \sqrt{k_{\cin, n}} v_{ne} G^+_n(\vect{r}),
\end{equation}
where $G^+_n(\vect{r})=\bra{\vect{r}}\hat{G}^+\ket{x_\cin, \chi_{\cin,n}}$.

\subsection{Radiance of transmission eigenchannels}\label{app:theory-radiance}

\par In the following, we are particularly interested in the mean intensity profile of transmission eigenchannels with a given transmission eigenvalue $T$:
\begin{equation}\label{eq:def-itprofile}
I_{T}(\vect{r}) = \frac{\avg{ \sum_{e=1}^{N_{\rm e}} \abs{\psi_{T_e}(\vect{r})}^2 \delta(T-T_e) }}{N_{\rm e}\rho(T)},
\end{equation}
where $\avg{\cdots}$ denotes the average over the disorder, $N_{\rm e} = \min(N_\cin,N_\cout)$ is the number of transmission eigenvalues, and $\rho(T) = \frac{1}{N_{\rm e}} \sum_{e=1}^{N_{\rm e}} \avg{\delta(T-T_e)}$ is the distribution of transmission eigenvalues.
Similarly, we define the average current density of transmission eigenchannels:
\begin{equation}\label{eq:def-jtcurrent}
\vect{j}_{T}(\vect{r}) = \frac{\avg{ \sum_{e=1}^{N_{\rm e}} \Im[ \cc{\psi}_{T_e}(\vect{r}) \grad_{\vect{r}}\psi_{T_e}(\vect{r}) ] \delta(T-T_e) }}{kN_{\rm e}\rho(T)} .
\end{equation}
More generally, we introduce the two-point local density of transmission eigenchannels:
\begin{equation}\label{eq:def-dote}
D_T(\vect{r},\vect{r}') = \frac{\avg{ \sum_{e=1}^{N_{\rm e}} \psi_{T_e}(\vect{r}) \cc{\psi}_{T_e}(\vect{r}') \delta(T-T_e) }}{N_{\rm e}\rho(T)} .
\end{equation}
This allows us to define the radiance of transmission eigenchannels as
\begin{equation}\label{eq:tstate-radiance}
f_T(\vect{u},\vect{r}) = \dashint_{k^2} \D{(p^2)} \nu(p) \int_{\mathbb{R}^d} \D{\vect{s}} \E^{-\I p\vect{u}\cdot\vect{s}} D_T(\vect{r}+\tfrac{\vect{s}}{2}, \vect{r}-\tfrac{\vect{s}}{2}).
\end{equation}
The intensity profile is then obtained from the directional average of the radiance,
\begin{equation}\label{eq:itprofile-from-radiance}
I_T(\vect{r}) = D_T(\vect{r},\vect{r}) = \oint \frac{\D{\vect{u}}}{S_d} f_T(\vect{u},\vect{r}) ,
\end{equation}
and the current density from
\begin{equation}\label{eq:jtcurrent-from-radiance}
\vect{j}_T(\vect{r}) = \oint \frac{\D{\vect{u}}}{S_d} \vect{u}\, f_T(\vect{u},\vect{r}) .
\end{equation}

\par To evaluate $D_T(\vect{r},\vect{r}')$, which completely determines the sought radiance $f_T(\vect{u},\vect{r})$ and intensity $I_T(\vect{r})$, we rewrite the transmission eigenchannel $\psi_{T_e}(\vect{r})$ of Eq.\ \eqref{eq:field-eigenchannel} in operator form,
\begin{equation}\label{eq:tstate-formal}
\psi_{T_e}(\vect{r}) = \sqrt{2}\I\mathcal{N} \sum_{n=1}^{N_{\cin}} v_{ne} \bra{\vect{r}} \op{G}^+ \op{K}_\cin^{\frac{1}{2}} \ket{x_\cin,\chi_{\cin,n}} ,
\end{equation}
where $\op{K}_\cin$ is the current density operator on the input surface \cite{GaspardD2025a, GaspardD2025b}
\begin{equation}\label{eq:def-contact-ka}
\op{K}_\cin = \sum_{n=1}^{N_\cin} 2k_{\cin,n} \ket{x_\cin,\chi_{\cin,n}} \bra{x_\cin,\chi_{\cin,n}} .
\end{equation}
Substituting Eq.\ \eqref{eq:tstate-formal} into Eq.\ \eqref{eq:def-dote} gives
\begin{equation}\label{eq:local-dote-formal}\begin{aligned}
D_T(\vect{r},\vect{r}') = \frac{2\abs{\mathcal{N}}^2}{N_{\rm e}\rho(T)} \avg{ \bra{\vect{r}} \op{G}^+ \op{K}_\cin^{\frac{1}{2}} \delta(T - \herm{\op{t}}\op{t}) \op{K}_\cin^{\frac{1}{2}} \op{G}^- \ket{\vect{r}'} } ,
\end{aligned}\end{equation}
where $\op{t}$ is the formal transmission operator defined in the full volume $\mathbb{R}^d$ by
\begin{equation}\label{eq:def-transmission-operator}
\op{t} = \I\op{K}_\cout^{\frac{1}{2}} \op{G}^+ \op{K}_\cin^{\frac{1}{2}} \:.
\end{equation}

\par The next step in evaluating $D_T(\vect{r}, \vect{r}')$ in Eq.\ \eqref{eq:local-dote-formal} is to introduce the $2\times 2$ matrix Green's function $\matr{\mathcal{G}}(\vect{r},\vect{r}')$, defined by \cite{GaspardD2025a, GaspardD2025b}
\begin{equation}\label{eq:def-matrix-green-1}\begin{aligned}
& \Big( \lapl_{\vect{r}} + k^2 + \I\varepsilon\matr{\sigma}_3 - U(\vect{r})  \\
& + \gamma_\cin\op{K}_\cin\matr{\sigma}_+ + \gamma_\cout\op{K}_\cout\matr{\sigma}_- \Big) \matr{\mathcal{G}}(\vect{r},\vect{r}') = \matr{1}_2\delta(\vect{r}-\vect{r}')  \:,
\end{aligned}\end{equation}
where $\matr{\sigma}_3=(\begin{smallmatrix}1 & 0\\ 0 & -1\end{smallmatrix})$, $\matr{\sigma}_+=(\begin{smallmatrix}0 & 1\\ 0 & 0\end{smallmatrix})$, and $\matr{\sigma}_-=(\begin{smallmatrix}0 & 0\\ 1 & 0\end{smallmatrix})$ are the Pauli matrices, $\op{K}_\cin$ and $\op{K}_\cout$ are the contact operators \eqref{eq:def-contact-ka}, and $\gamma_\cin$ and $\gamma_\cout$ are coupling parameters that depend on the transmission eigenvalue $T$ according to
\begin{equation}\label{eq:def-gamma}
\gamma_\cin = \gamma_\cout = \frac{1}{\sqrt{T}} + 0^+\I .
\end{equation}
Using the inversion formula for $2\times 2$ block matrices, the upper-right element of $\op{\matr{\mathcal{G}}}$ can be expressed in terms of the transmission operator.
Substituting Eq.\ \eqref{eq:def-transmission-operator} and taking the imaginary part gives
\begin{equation}\label{eq:im-g12-op}
\Im \op{\mathcal{G}}_{12} = \frac{-\pi}{\gamma_\cout} \op{G}^+ \op{K}_\cin^{\frac{1}{2}} \delta(T - \herm{\op{t}} \op{t}) \op{K}_\cin^{\frac{1}{2}} \op{G}^-  \:.
\end{equation}
Combining this relation with Eq.\ \eqref{eq:local-dote-formal}, we obtain
\begin{equation}\label{eq:dote-from-im-g12}
D_T(\vect{r},\vect{r}') = -\frac{2\abs{\mathcal{N}}^2 \gamma_\cout}{\pi N_{\rm e}\rho(T)} \Im\avg{\mathcal{G}_{12}(\vect{r},\vect{r}')}.
\end{equation}

\par The remaining task is therefore to determine the disorder-averaged matrix Green's function.
Following the replica-method derivation developed in Refs.\ \cite{GaspardD2025a, GaspardD2025b}, the disorder average leads to a nonlinear equation for $\avg{\matr{\mathcal{G}}(\vect{r},\vect{r}')}$,
\begin{equation}\label{eq:matrix-green-avg}\begin{aligned}
& \Big( \lapl_{\vect{r}} + k^2 + \I\varepsilon\matr{\sigma}_3 - \tfrac{k}{\ell\pi\nu} \avg{\matr{\mathcal{G}}(\vect{r},\vect{r})} \\
& + \gamma_\cin\op{K}_\cin\matr{\sigma}_+ + \gamma_\cout\op{K}_\cout\matr{\sigma}_- \Big) \avg{\matr{\mathcal{G}}(\vect{r},\vect{r}')} = \matr{1}_2\delta(\vect{r}-\vect{r}') ,
\end{aligned}\end{equation}
where $\ell$ is the mean free path.
Equation \eqref{eq:matrix-green-avg} has the same structure as Eq.\ \eqref{eq:def-matrix-green-1}, except that the scalar random potential $U(\vect{r})$ is replaced by the matrix self-energy $\tfrac{k}{\ell\pi\nu} \avg{\matr{\mathcal{G}}(\vect{r},\vect{r})}$.
This self-energy depends on the Green's function itself, making the resulting equation nonlinear.

\par We next introduce the matrix radiance $\matr{g}(\vect{u},\vect{r})$ through the Wigner transform
\begin{equation}\label{eq:def-matrix-radiance}
\matr{g}(\vect{u},\vect{r}) = \frac{\I}{\pi} \dashint_{k^2} \D{(p^2)} \int \D{\vect{s}}\,\E^{-\I p\vect{u}\cdot\vect{s}} \avg{\matr{\mathcal{G}}(\vect{r}+\tfrac{\vect{s}}{2}, \vect{r}-\tfrac{\vect{s}}{2})} .
\end{equation}
The directional moments of this matrix radiance define the matrix field
\begin{equation}\label{eq:qn-from-g}
\matr{Q}(\vect{r}) = \oint \frac{\D{\vect{u}}}{S_d} \matr{g}(\vect{u},\vect{r}) ,
\end{equation}
and the matrix current
\begin{equation}\label{eq:jn-from-g}
\vect{\matr{J}}(\vect{r}) = \oint \frac{\D{\vect{u}}}{S_d} \vect{u}\,\matr{g}(\vect{u},\vect{r}) ,
\end{equation}
on an equal footing with Eqs.\ \eqref{eq:itprofile-from-radiance} and \eqref{eq:jtcurrent-from-radiance}, respectively.
Under the semiclassical approximation, Eq.\ \eqref{eq:matrix-green-avg}, restricted to the domain of the scattering medium, takes the form of a matrix transport equation,
\begin{equation}\label{eq:eilenberger-short}
\vect{u}\cdot\grad_{\vect{r}}\matr{g}(\vect{u},\vect{r}) = -\tfrac{1}{2\ell} [\matr{Q}(\vect{r}), \matr{g}(\vect{u},\vect{r})] - \tfrac{1}{2\labso} [\matr{\sigma}_3, \matr{g}(\vect{u},\vect{r})] ,
\end{equation}
which is Eq.\ (2) of the main text.
The contact terms in Eq.\ \eqref{eq:matrix-green-avg} do not enter the bulk transport equation; their effect is instead incorporated through the appropriate boundary conditions \cite{GaspardD2025b}.
The matrix radiance $\matr{g}(\vect{u},\vect{r})$ is subject to the constraint 
\begin{equation}\label{eq:g2-equals-one}
\matr{g}(\vect{u},\vect{r})^2 = \matr{1}_2 ,\qquad\forall\,\vect{r},\vect{u} ,
\end{equation}
which is due to the nonlinearity of Eq.\ \eqref{eq:matrix-green-avg}.
See also Refs.\ \cite{GaspardD2025a, GaspardD2025b} for details of the derivation of Eq.\ \eqref{eq:eilenberger-short}.

\par We can now establish the connection between the matrix radiance and the radiance of transmission eigenchannels.
Combining Eqs.\ \eqref{eq:tstate-radiance}, \eqref{eq:dote-from-im-g12}, and \eqref{eq:def-matrix-radiance}, and using the normalization factor \eqref{eq:normalization-prefactor}, gives
\begin{equation}\label{eq:tradiance-from-g}\begin{aligned}
f_T(\vect{u},\vect{r}) & = \frac{2\abs{\mathcal{N}}^2\nu}{N_{\rm e}\rho(T)\sqrt{T}} \Re g_{12}(\vect{u},\vect{r}) \\
 & = \frac{N_{\cin}}{\pi N_{\rm e}\rho(T)\sqrt{T}} \Re g_{12}(\vect{u},\vect{r}) ,
\end{aligned}\end{equation}
in agreement with the formula given in the Methods section, since $N_\cin/N_{\rm e}=S_{\cin}/S_{\rm e}$.

\par Finally, combining Eq.\ \eqref{eq:tradiance-from-g} with Eqs.\ \eqref{eq:itprofile-from-radiance} and \eqref{eq:qn-from-g} yields the intensity profile of transmission eigenchannels
\begin{equation}\label{eq:itprofile-from-qn}
I_{T}(\vect{r}) = \frac{N_\cin}{\pi N_{\rm e}\rho(T)\sqrt{T}} \Re Q_{12}(\vect{r}) ,
\end{equation}
and with Eqs.\ \eqref{eq:jtcurrent-from-radiance} and \eqref{eq:jn-from-g}, the current density of transmission eigenchannels
\begin{equation}\label{eq:jtcurrent-from-jn}
\vect{j}_{T}(\vect{r}) = \frac{N_\cin}{\pi N_{\rm e}\rho(T)\sqrt{T}} \Re\vect{J}_{12}(\vect{r}) .
\end{equation}

\subsection{Transmission eigenvalue distribution}\label{app:theory-distribution}

\par The calculation above also provides the distribution of transmission eigenvalues $\rho(T)$.
Indeed, combining Eqs.\ \eqref{eq:def-transmission-operator} and \eqref{eq:im-g12-op} gives
\begin{equation}\label{eq:im-g12-kout}
\Im\Tr\left( \tavg{\op{\mathcal{G}}_{12}}\op{K}_\cout \right) = \frac{-\pi}{\gamma_\cout} \avg{\Tr(\herm{\op{t}} \op{t} \delta(T - \herm{\op{t}} \op{t}))},
\end{equation}
so that
\begin{equation}\label{eq:tspectrum-from-g12}
\rho(T) = \frac{-\gamma_\cout}{\pi N_{\rm e}T} \Im\Tr\big( \tavg{\op{\mathcal{G}}_{12}}\op{K}_\cout \big) .
\end{equation}

\par To express the remaining trace in Eq.\ \eqref{eq:tspectrum-from-g12} in terms of the matrix current, we use the fact that, according to Eq.\ \eqref{eq:def-contact-ka}, $\op{K}_\cout$ acts as a current density operator on the output surface.
This gives
\begin{equation}\label{eq:trace-g12-kb-from-jn}
\Tr\big(\tavg{\op{\matr{\mathcal{G}}}}\op{K}_\cout\big) = -2\I\pi\nu k \int_{\mathcal{S}_\cout} \D{\vect{y}}\cdot\vect{\matr{J}}(\vect{r}) ,
\end{equation}
where $\vect{\matr{J}}(\vect{r})$ is the dimensionless matrix current \eqref{eq:jn-from-g}, which can equivalently be expressed in terms of the disorder-averaged matrix Green's function as
\begin{equation}\label{eq:jn-from-green}
\vect{\matr{J}}(\vect{r}) = \frac{\I}{\pi\nu k} \lim_{\vect{r}'\rightarrow\vect{r}} \frac{\grad_{\vect{r}}\avg{\matr{\mathcal{G}}(\vect{r},\vect{r}')} - \grad_{\vect{r}'}\avg{\matr{\mathcal{G}}(\vect{r},\vect{r}')}}{2\I}  \:.
\end{equation}
Substituting Eq.\ \eqref{eq:trace-g12-kb-from-jn} into Eq.\ \eqref{eq:tspectrum-from-g12} then gives
\begin{equation}\label{eq:tspectrum-from-jn-gen}
\rho(T) = \frac{2\pi\nu k\gamma_\cout}{\pi N_{\rm e}T} \Re \int_{\mathcal{S}_\cout} \D{\vect{y}}\cdot\vect{J}_{12}(\vect{r}) .
\end{equation}

\par The prefactor can be simplified by using the relation between the surface density of waveguide modes and the bulk density of states $\nu$,
\begin{equation}\label{eq:mode-density}
\frac{N_{\rm e}}{S_{\rm e}} = \frac{V_{d-1}k^{d-1}}{(2\pi)^{d-1}} = \frac{2\pi\nu k}{d\mu} ,
\end{equation}
where $S_{\rm e}=\min(S_\cin, S_\cout)$, $\mu$ is the mean direction cosine given by
\begin{equation}\label{eq:mean-dir-cos-k1}
\mu = \frac{V_d}{2V_{d-1}} = \begin{cases}
1              & (d=1) ,\\
\tfrac{\pi}{4} & (d=2) ,\\
\tfrac{2}{3}   & (d=3) ,
\end{cases}\end{equation}
and $V_d$ is the volume of the unit $d$-ball in $\mathbb{R}^d$.
Using Eqs.\ \eqref{eq:def-gamma} and \eqref{eq:mode-density}, Eq.\ \eqref{eq:tspectrum-from-jn-gen} can finally be written as
\begin{equation}\label{eq:tspectrum-from-jn}
\rho(T) = \frac{d\mu}{\pi S_{\rm e}T^{\frac{3}{2}}} \Re \int_{\mathcal{S}_\cout} \D{\vect{y}}\cdot\vect{J}_{12}(\vect{r}).
\end{equation}

\section{Diffusion approximation}\label{app:diffusion-approx}

\subsection{Matrix diffusion equation}\label{app:usadel-derivation}

\par Let us derive the matrix diffusion equation from the matrix transport equation \eqref{eq:eilenberger-short}.
Integrating Eq.\ \eqref{eq:eilenberger-short} over the propagation directions gives the continuity equation
\begin{equation}\label{eq:usadel-mom0}
\grad_{\vect{r}}\cdot\vect{\matr{J}}(\vect{r}) = -\tfrac{1}{2\labso} [\matr{\sigma}_3, \matr{Q}(\vect{r})] ,
\end{equation}
while multiplying it by $\vect{u}$ and integrating over directions gives the matrix analogue of Fick's law,
\begin{equation}\label{eq:usadel-mom1}
\tfrac{1}{d}\grad_{\vect{r}}\matr{Q}(\vect{r}) = \tfrac{1}{2\ell} [\vect{\matr{J}}(\vect{r}), \matr{Q}(\vect{r})] - \tfrac{1}{2\labso} [\matr{\sigma}_3, \vect{\matr{J}}(\vect{r})] ,
\end{equation}
In the diffusive regime, the directional dependence of the radiance $\matr{g}(\vect{u},\vect{r})$ is weak, so that it can be expanded to first order in $\vect{u}$:
\begin{equation}\label{eq:g-radiance-p1-expansion}
\matr{g}(\vect{u},\vect{r}) = \matr{Q}(\vect{r}) + d\vect{u}\cdot\vect{\matr{J}}(\vect{r}) + \bigo(\vect{u}^2) .
\end{equation}
This expansion must remain consistent with the nonlinear constraint $\matr{g}(\vect{u},\vect{r})^2=\matr{1}_2$.
To first order in the directional expansion, this requires
\begin{equation}\label{eq:g-radiance-p1-condition}\begin{split}
\matr{g}(\vect{u},\vect{r})^2 & = \left( \matr{Q}(\vect{r}) + d\vect{u}\cdot\vect{\matr{J}}(\vect{r}) + \bigo(\vect{u}^2) \right)^2  \:,\\
 & = \matr{Q}(\vect{r})^2 + d\vect{u}\cdot\{\matr{Q}(\vect{r}), \vect{\matr{J}}(\vect{r})\} + \bigo(\vect{u}^2)  \:,\\
 & = \matr{1}_2  \:,
\end{split}\end{equation}
where $\{\matr{A}, \matr{B}\}=\matr{A}\matr{B}+\matr{B}\matr{A}$ is the matrix anticommutator.
Equating the different orders in $\vect{u}$ then gives the constraints
\begin{equation}\label{eq:usadel-q2-qj}
\matr{Q}(\vect{r})^2 = \matr{1}_2  \:,\qquad  \{\matr{Q}(\vect{r}), \vect{\matr{J}}(\vect{r})\} = \matr{0}  \:.
\end{equation}
At this order, we neglect terms of order $\bigo(\vect{u}^2)$ and higher, including $[\vect{u}\cdot\vect{\matr{J}}(\vect{r})]^2$.
The resulting constraints are therefore an approximation to the exact normalization $\matr{g}(\vect{u},\vect{r})^2 = \matr{1}_2$, since $\matr{Q}(\vect{r})$ is itself the directional average of $\matr{g}(\vect{u},\vect{r})$.

\par Furthermore, the absorption term in Eq.\ \eqref{eq:usadel-mom1} must be neglected in order to preserve the constraint $\matr{Q}(\vect{r})^2=\matr{1}_2$.
Indeed, taking the anticommutator of Eq.\ \eqref{eq:usadel-mom1} with $\matr{Q}(\vect{r})$ and temporarily omitting the positional dependence gives
\begin{equation}\label{eq:usadel-mom1-anticom-q-1}
\tfrac{1}{d}\{\matr{Q}, \grad_{\vect{r}}\matr{Q}\} = -\tfrac{1}{2\ell} \{\matr{Q}, [\matr{Q}, \vect{\matr{J}}]\} - \tfrac{1}{2\labso} \{\matr{Q}, [\matr{\sigma}_3, \vect{\matr{J}}]\} \:,
\end{equation}
which, using the facts that $\grad_{\vect{r}}(\matr{Q}^2) = \{\matr{Q}, \grad_{\vect{r}}\matr{Q}\}$ and $\{\matr{Q}, [\matr{Q}, \matr{A}]\} = [\matr{Q}^2, \matr{A}] = [\matr{1}_2, \matr{A}] = \matr{0}$, reduces to
\begin{equation}\label{eq:usadel-mom1-anticom-q-2}
\tfrac{1}{d} \grad_{\vect{r}}(\matr{Q}^2) = -\tfrac{1}{2\labso} \{\matr{Q}, [\matr{\sigma}_3, \vect{\matr{J}}]\} .
\end{equation}
Thus, absorption would cause $\matr{Q}(\vect{r})^2$ to deviate from $\matr{1}_2$.
Since we consider the weak-absorption regime $\labso^{-1}\ll\ell^{-1}$, we neglect this term at the diffusive level.
The matrix diffusion equation then takes the form
\begin{equation}\label{eq:usadel-1st-form}\begin{aligned}
\grad_{\vect{r}}\cdot\vect{\matr{J}}(\vect{r}) & = -\tfrac{1}{2\labso} [\matr{\sigma}_3, \matr{Q}(\vect{r})] ,\\
\tfrac{1}{d}\grad_{\vect{r}}\matr{Q}(\vect{r}) & = -\tfrac{1}{2\ell} [\matr{Q}(\vect{r}), \vect{\matr{J}}(\vect{r})] .
\end{aligned}\end{equation}

\par Using the anticommutation relation in Eq.\ \eqref{eq:usadel-q2-qj}, the second equation of Eq.\ \eqref{eq:usadel-1st-form} can be solved for the matrix current, giving
\begin{equation}\label{eq:usadel-2nd-form}\begin{aligned}
\grad_{\vect{r}}\cdot\vect{\matr{J}}(\vect{r}) & = -\tfrac{1}{2\labso} [\matr{\sigma}_3, \matr{Q}(\vect{r})] ,\\
\vect{\matr{J}}(\vect{r}) & = -\tfrac{\ell}{d} \matr{Q}(\vect{r}) \grad_{\vect{r}}\matr{Q}(\vect{r}) ,
\end{aligned}\end{equation}
in accordance with Eq.\ (3) of the main text.
Equation \eqref{eq:usadel-2nd-form} is nothing but a closed equation for $\matr{Q}(\vect{r})$:
\begin{equation}\label{eq:usadel-3rd-form}
\grad_{\vect{r}}\cdot\big( \matr{Q}(\vect{r}) \grad_{\vect{r}}\matr{Q}(\vect{r}) \big) = \tfrac{d}{2\ell\labso} [\matr{\sigma}_3, \matr{Q}(\vect{r})] .
\end{equation}
This equation is analogous to the Usadel equation of the superconductivity literature \cite{Usadel1970, Kamenev2023} and can also be obtained as the saddle-point equation of a nonlinear sigma model Lagrangian (see Eq.\ (A38) of Ref.\ \cite{GaspardD2025a}).

\subsection{Boundary conditions of the matrix diffusion equation}\label{app:usadel-boundary}%

\par To solve the matrix diffusion equation \eqref{eq:usadel-2nd-form} for $\matr{Q}(\vect{r})$ in the bulk of the disordered region, we need to supplement it with appropriate boundary conditions.
We distinguish between free (uncontrolled) surfaces, through which the wave can escape, and the input and output interfaces, where the wave is injected or detected.
The latter require additional boundary conditions arising from the contact terms in Eq.\ \eqref{eq:matrix-green-avg}.

\par According to Eq.\ (99) of Ref.\ \cite{GaspardD2025b}, the contact terms modify the value of $\matr{Q}$ across the input and output interfaces.
More precisely, the values immediately inside and outside the scattering region are related by
\begin{equation}\label{eq:qn-inout-contacts}\begin{aligned}
\matr{Q}(\vect{r}_\cin^+)  & = \E^{\frac{\I}{\sqrt{T}} \matr{\sigma}_+} \matr{Q}(\vect{r}_\cin^-)  \E^{-\frac{\I}{\sqrt{T}} \matr{\sigma}_+}  \:,\\
\matr{Q}(\vect{r}_\cout^-) & = \E^{-\frac{\I}{\sqrt{T}} \matr{\sigma}_-} \matr{Q}(\vect{r}_\cout^+) \E^{\frac{\I}{\sqrt{T}} \matr{\sigma}_-}  \:,
\end{aligned}\end{equation}
where $\vect{r}_\cin^+$ ($\vect{r}_\cout^-$) and $\vect{r}_\cin^-$ ($\vect{r}_\cout^+$) denote points immediately inside and outside the input (output) surface, respectively.
Here, $T$ is the transmission eigenvalue, and $\matr{\sigma}_\pm=(\matr{\sigma}_1\pm\I\matr{\sigma}_2)/2$ are the ladder Pauli matrices.

\par It remains to determine the values of $\matr{Q}$ on the exterior side of these interfaces.
These can be obtained from the boundary conditions for the matrix radiance $\matr{g}(\vect{u},\vect{r})$ in the exterior of the scattering region, derived in Ref.\ \cite{GaspardD2025b}.
In the absence of specular reflection at the interface, these conditions read
\begin{equation}\label{eq:eilenberger-bdr-1}
\matr{g}(\vect{u},\vect{r}) = \begin{cases}
\matr{g}_\cin(\vect{u},\vect{r})   & (\vect{u}\cdot\vect{n} < 0) ,\\
\matr{g}_\cout(\vect{u},\vect{r})  & (\vect{u}\cdot\vect{n} > 0),
\end{cases}\end{equation}
where $\vect{n}$ is a unit vector pointing outward from the scattering region at the interface $\partial\mathcal{V}$.
The matrices $\matr{g}_\cin(\vect{u},\vect{r})$ and $\matr{g}_\cout(\vect{u},\vect{r})$ describing incoming and outgoing directions are
\begin{equation}\label{eq:eilenberger-bdr-2}\begin{aligned}
& \matr{g}_\cin(\vect{u},\vect{r})  = \begin{pmatrix}1 & 0\\ g_{21}(\vect{u},\vect{r}) & -1\end{pmatrix} ,\\
& \matr{g}_\cout(\vect{u},\vect{r}) = \begin{pmatrix}1 & g_{12}(\vect{u},\vect{r})\\ 0 & -1\end{pmatrix} ,
\end{aligned}\end{equation}
where $g_{12}(\vect{u},\vect{r})$ and $g_{21}(\vect{u},\vect{r})$ are not fixed by the boundary conditions.

\par To obtain boundary conditions consistent with the diffusive approximation \eqref{eq:g-radiance-p1-expansion}, we take the directional average of Eq.\ \eqref{eq:eilenberger-bdr-1}:
\begin{equation}\label{eq:eilenberger-bdr-avg}\begin{aligned}
\avg{\matr{g}(\vect{u},\vect{r})}^+ & = \avg{\matr{g}_\cin(\vect{u},\vect{r})}^+ ,\\
\avg{\matr{g}(\vect{u},\vect{r})}^- & = \avg{\matr{g}_\cout(\vect{u},\vect{r})}^- ,
\end{aligned}\end{equation}
where 
\begin{equation}
\avg{\matr{g}(\vect{u},\vect{r})}^\pm = \frac{2\pi}{S_{d+1}} \int_{\pm u>0} \D{\vect{u}} \abs{u} \matr{g}(\vect{u},\vect{r}) .
\end{equation}
Using the Jacobian $\D{\vect{u}}=S_{d-1}(1-u^2)^{\frac{d-3}{2}}\D{u}$ and the expansion \eqref{eq:g-radiance-p1-expansion}, we obtain
\begin{equation}
\avg{\matr{g}(\vect{u},\vect{r})}^\pm = \matr{Q}(\vect{r}) \pm d \mu\vect{n}\cdot\vect{\matr{J}}(\vect{r}) ,
\end{equation}
where $\mu$ is given by Eq.\ \eqref{eq:mean-dir-cos-k1}.
The boundary conditions \eqref{eq:eilenberger-bdr-avg} therefore give
\begin{equation}\label{eq:usadel-bdr-qn-jn-1}\begin{aligned}
\matr{Q}(\vect{r}) & = \begin{pmatrix}1 & \tfrac{1}{2}\avg{g_{12}}^-\\ \tfrac{1}{2}\avg{g_{21}}^+ & -1\end{pmatrix} ,\\
d\mu \vect{n}\cdot\vect{\matr{J}}(\vect{r}) & = \begin{pmatrix}0 & -\tfrac{1}{2}\avg{g_{12}}^-\\ \tfrac{1}{2}\avg{g_{21}}^+ & 0\end{pmatrix} .
\end{aligned}\end{equation}
Eliminating the unspecified matrix elements yields the boundary condition
\begin{equation}\label{eq:usadel-boundary-1}
[\matr{\sigma}_3,\matr{Q}(\vect{r})] = 2d\mu \vect{n}\cdot\vect{\matr{J}}(\vect{r})  \:,\quad\forall\vect{r}\in\partial\mathcal{V}  \:.
\end{equation}
A similar boundary condition was obtained in Eq.\ (40) of Ref.\ \cite{TianC2008b}.

\par Using $\vect{\matr{J}}(\vect{r}) = -\tfrac{\ell}{2d} [\matr{Q}(\vect{r}), \grad_{\vect{r}}\matr{Q}(\vect{r})]$ and approximating the normal derivative by
\begin{equation}
\vect{n}\cdot\grad_{\vect{r}}\matr{Q}(\vect{r}) \simeq \frac{\matr{Q}(\vect{r}+z_0\vect{n}) - \matr{Q}(\vect{r})}{z_0},
\end{equation}
where $z_0=\mu \ell$ is the extrapolation length familiar from classical diffusion, we obtain
\begin{equation}\label{eq:jn-normal-from-extrapol}
\vect{n}\cdot\vect{\matr{J}}(\vect{r}) = \tfrac{1}{2d\mu} \left[ \matr{Q}(\vect{r}+z_0\vect{n}), \matr{Q}(\vect{r}) \right]  \:.
\end{equation}
Substitution into Eq.\ \eqref{eq:usadel-boundary-1} then gives the simpler boundary condition
\begin{equation}\label{eq:usadel-bdr-loss}
\matr{Q}(\vect{r}+z_0\vect{n}) = \matr{\sigma}_3  \:,\quad\forall\vect{r}\in\partial\mathcal{V}  \:,
\end{equation}
which states that $\matr{Q}$ extrapolated a distance $z_0$ outside the medium along the outward normal must equal $\matr{\sigma}_3$.

\par For a free surface, Eq.\ \eqref{eq:usadel-bdr-loss} directly provides the required boundary condition.
At the input and output interfaces, however, it applies to the exterior values $\matr{Q}(\vect{r}_\cin^-)$ and $\matr{Q}(\vect{r}_\cout^+)$.
Combining these conditions with the contact relations \eqref{eq:qn-inout-contacts}, we obtain the boundary conditions for the matrix diffusion equation:
\begin{equation}\label{eq:usadel-bdr-inout-1}\begin{aligned}
& \matr{Q}(\vect{r}_\cin  + z_0\vect{n}) = \matr{Q}_\cin ,\\
& \matr{Q}(\vect{r}_\cout + z_0\vect{n}) = \matr{Q}_\cout ,
\end{aligned}\end{equation}
where
\begin{equation}\label{eq:usadel-bdr-inout-2}\begin{aligned}
& \matr{Q}_\cin = \E^{\frac{\I}{\sqrt{T}} \matr{\sigma}_+} \matr{\sigma}_3 \E^{-\frac{\I}{\sqrt{T}} \matr{\sigma}_+} = \begin{pmatrix}1 & \frac{-2\I}{\sqrt{T}}\\ 0 & -1\end{pmatrix} ,\\
& \matr{Q}_\cout = \E^{-\frac{\I}{\sqrt{T}} \matr{\sigma}_-} \matr{\sigma}_3 \E^{+\frac{\I}{\sqrt{T}} \matr{\sigma}_-} = \begin{pmatrix}1 & 0\\ \frac{-2\I}{\sqrt{T}} & -1\end{pmatrix}.
\end{aligned}\end{equation}

\par The boundary conditions given above apply to non-reflecting boundaries opening onto the space outside the disordered medium.
In the presence of reflecting interfaces, however, different boundary conditions apply, as shown in Refs.\ \cite{Zaitsev1984, Nazarov1999b}.
In the specific case of perfectly reflecting walls considered in the present work to model lossless media, these boundary conditions impose the vanishing of the matrix current in the direction normal to the wall:
\begin{equation}\label{eq:usadel-bdr-mirror-jn}
\vect{n}\cdot\vect{\matr{J}}(\vect{r}) = 0 .
\end{equation}
In terms of $\matr{Q}(\vect{r})$, this translates to
\begin{equation}\label{eq:usadel-bdr-mirror-qn}
\vect{n}\cdot\grad_{\vect{r}}\matr{Q}(\vect{r}) = 0 .
\end{equation}

\section{Analytical solution in a lossless disordered medium}\label{app:lossless-solution}%

\par In this Appendix, we derive an analytical solution of the matrix diffusion equation \eqref{eq:usadel-2nd-form} for a lossless diffusive system of arbitrary structure.
We then use this solution to obtain the transmission eigenvalue distribution, as well as the intensity profile and current density of transmission eigenchannels.
Finally, we compare the latter with the classical diffusive solution and show why transmission eigenchannel profiles cannot, in general, be interpreted as eigenmodes of a diffusion equation.

\subsection{Solution of the matrix diffusion equation}\label{app:lossless-solution-matrix}

\par The constraint $\matr{Q}(\vect{r})^2 = \matr{1}_2$ allows us to represent the traceless matrix $\matr{Q}(\vect{r})$ as a vector on a (complexified) two-sphere in the three-dimensional space of Pauli matrices.
The two boundary values $\matr{Q}_\cin$ and $\matr{Q}_\cout$ therefore define a great circle on this sphere.
Its normal is determined by their commutator, which motivates the definition
\begin{equation}\label{eq:def-k2-axis}
\matr{K}_2 = \frac{[\matr{Q}_\cout, \matr{Q}_\cin]}{\sqrt{[\matr{Q}_\cout, \matr{Q}_\cin]^2}} 
 = \frac{1}{\sqrt{1-T}} \begin{pmatrix}1 & -\I\sqrt{T}\\ -\I\sqrt{T} & -1\end{pmatrix} .
\end{equation}
The notation $\matr{K}_2$ anticipates the geometric representation presented in Appendix \ref{app:spherical-transport}.
Using the boundary values given in Eq.\ \eqref{eq:usadel-bdr-inout-2}, one obtains the second equality in Eq.\ \eqref{eq:def-k2-axis}.

\par As discussed below, in the absence of both absorption and free surfaces, the matrix diffusion equation admits a solution confined to this great circle.
It can therefore be parametrized by a single angular coordinate $\theta(\vect{r})$,
\begin{equation}\label{eq:qn-param-lossless}
\matr{Q}(\vect{r}) = \E^{-\frac{\I}{2}\matr{K}_2\theta(\vect{r})} \matr{Q}_\cout \E^{\frac{\I}{2}\matr{K}_2\theta(\vect{r})} .
\end{equation}
The angle $\theta(\vect{r})$ measures the angular position of $\matr{Q}(\vect{r})$ along this great circle relative to $\matr{Q}_\cout$, with
\begin{equation}\label{eq:costheta-from-qn}
\cos\theta(\vect{r}) = \tfrac{1}{2}\Tr[\matr{Q}(\vect{r})\matr{Q}_\cout] .
\end{equation}

\par Since $\matr{K}_2$ is constant and anticommutes with $\matr{Q}(\vect{r})$ at all positions $\vect{r}$, the corresponding matrix current is
\begin{equation}\label{eq:jn-from-grad-theta}
\vect{\matr{J}}(\vect{r}) = -\frac{\ell}{d} \matr{Q}(\vect{r}) \grad_{\vect{r}}\matr{Q}(\vect{r}) = -\frac{\I\ell}{d}\matr{K}_2\grad_{\vect{r}}\theta(\vect{r}) .
\end{equation}
Thus, all matrix elements of the current share the same target-space direction $\matr{K}_2$, while their spatial dependence is carried by $\grad_{\vect{r}}\theta(\vect{r})$.
Current conservation therefore reduces to
\begin{equation}\label{eq:usadel-kappa-lossless}
\lapl_{\vect{r}}\theta(\vect{r}) = 0 .
\end{equation}
The boundary conditions for $\theta(\vect{r})$ follow from Eqs.\ \eqref{eq:usadel-bdr-inout-1}--\eqref{eq:usadel-bdr-inout-2} and read
\begin{equation}\label{eq:theta-bdr}\begin{aligned}
\theta(\vect{r}_\cin  + z_0\vect{n}) & = \theta_\cin ,\\
\theta(\vect{r}_\cout + z_0\vect{n}) & = 0,
\end{aligned}\end{equation}
where
\begin{equation}\label{eq:costheta-input}
\cos(\theta_\cin) = \tfrac{1}{2}\Tr[\matr{Q}_\cin\matr{Q}_\cout].
\end{equation}
Using Eq.\ \eqref{eq:usadel-bdr-inout-2}, this gives
\begin{equation}\label{eq:theta-input}
\theta_\cin = \pi + 2\I\arccosh(\tfrac{1}{\sqrt{T}}) .
\end{equation}
Equations \eqref{eq:qn-param-lossless}, \eqref{eq:def-k2-axis}, \eqref{eq:usadel-kappa-lossless}, and \eqref{eq:theta-bdr} therefore provide the complete solution for $\matr{Q}(\vect{r})$.

\subsection{Connection with the classical diffusive solution}\label{app:lossless-solution-connection}

\par The scalar equation \eqref{eq:usadel-kappa-lossless} has the same form as the diffusion equation governing the intensity associated with a random incident wavefront.
This provides a direct connection between the matrix and classical diffusion problems.
We define
\begin{equation}\label{eq:ibar-from-theta-alt}
\bar{I}(\vect{r}) = 2\frac{\theta(\vect{r})}{\theta_\cin} .
\end{equation}
The intensity $\bar{I}(\vect{r})$ then satisfies
\begin{equation}\label{eq:ibar-diffusion}\begin{aligned}
\lapl_{\vect{r}}\bar{I}(\vect{r}) & = 0, \\
\bar{I}(\vect{r}_\cin  + z_0\vect{n}) & = 2 ,\\
\bar{I}(\vect{r}_\cout + z_0\vect{n}) & = 0 .
\end{aligned}\end{equation}
This is precisely the classical diffusion problem for a uniformly illuminated input surface.
Indeed, the corresponding intensity can be written as
\begin{equation}\label{eq:ibar-from-green}
\bar{I}(\vect{r}) = \frac{4\abs{\mathcal{N}}^2}{N_{\rm e}} \sum_{n=1}^{N_\cin} k_{\cin,n} \avg{ \abs{G^+_n(\vect{r})}^2 } ,
\end{equation}
which, according to the definitions \eqref{eq:field-eigenchannel} and \eqref{eq:def-itprofile}, is equivalently given by the incoherent sum of all transmission eigenchannel profiles,
\begin{equation}\label{eq:ibar-from-itprofile}
\bar{I}(\vect{r}) = \int_{0^+}^1 \D{T}\,\rho(T) I_{T}(\vect{r}).
\end{equation}
The factor of 2 in the second line of Eq.\ \eqref{eq:ibar-diffusion} arises because, in the diffusive regime, the intensity at the input consists of the incoherent superposition of the incident and reflected waves, both of which have unit intensity.

\par The current density associated with $\bar{I}(\vect{r})$ is
\begin{equation}\label{eq:jbar-from-ibar}
\bar{\vect{j}}(\vect{r}) = -\frac{\ell}{d} \grad_{\vect{r}}\bar{I}(\vect{r}) ,
\end{equation}
and its total transmitted flux gives the mean transmission,
\begin{equation}\label{eq:tavg-from-jbar}
\bar{T} = \frac{\int_{\mathcal{S}_\cout} \D{\vect{y}}\cdot\bar{\vect{j}}(\vect{r})}{\int_{\mathcal{S}_\cin} \D{\vect{y}}\cdot(-\vect{j}_{\cin}(\vect{r}))} = \frac{d\mu}{S_\cin} \int_{\mathcal{S}_\cout} \D{\vect{y}}\cdot\bar{\vect{j}}(\vect{r}),
\end{equation}
where $\vect{j}_{\cin}(\vect{r})$ is the incident current associated with a field normalized to unit intensity, $\int_{\mathcal{S}_\cin} \frac{\D{\vect{y}}}{S_\cin} \abs{\psi_\cin(\vect{r})}^2 = 1$.
This quantity is equal to the mean transmission probability, $\bar{T} = \frac{1}{N_{\rm e}} \avg{\Tr(\herm{\matr{t}}\matr{t})}$.

\par Expressed in terms of the classical diffusive solution, the matrix solution \eqref{eq:qn-param-lossless} becomes
\begin{equation}\label{eq:qn-param-lossless-2}
\matr{Q}(\vect{r}) = \E^{-\frac{\I\theta_\cin}{4}\matr{K}_2 \bar{I}(\vect{r})} \matr{Q}_\cout \E^{\frac{\I \theta_\cin}{4}\matr{K}_2 \bar{I}(\vect{r})} ,
\end{equation}
while the corresponding matrix current is
\begin{equation}\label{eq:jn-from-jbar}
\vect{\matr{J}}(\vect{r}) = -\tfrac{\ell}{d} \matr{Q}(\vect{r}) \grad_{\vect{r}}\matr{Q}(\vect{r}) = \frac{\I\theta_\cin}{2} \matr{K}_2 \bar{\vect{j}}(\vect{r}) .
\end{equation}
Thus, remarkably, the spatial dependence of the matrix solution is entirely determined by the classical diffusive solution $\bar{I}(\vect{r})$, while the dependence on the transmission eigenvalue $T$ is encoded in the matrix structure through $\theta_\cin$ and $\matr{K}_2$.

\subsection{Bimodal distribution in arbitrary structure}\label{app:bimodal-any-structure}%

\par We can now obtain the transmission eigenvalue distribution directly from the matrix current.
Substituting Eq.\ \eqref{eq:jn-from-jbar} into Eq.\ \eqref{eq:tspectrum-from-jn} gives
\begin{equation}\label{eq:tspectrum-from-jbar}
\rho(T) = \frac{1}{2T\sqrt{1-T}} \frac{d\mu}{S_{\rm e}} \int_{\mathcal{S}_\cout} \D{\vect{y}}\cdot\bar{\vect{j}}(\vect{r}) .
\end{equation}
Using Eq.\ \eqref{eq:tavg-from-jbar} and the proportionality $\frac{N_\cin}{S_\cin}=\frac{N_{\rm e}}{S_{\rm e}}$, we obtain
\begin{equation}\label{eq:bimodal-normalized}
\rho(T) = \frac{N_\cin}{N_{\rm e}} \frac{\bar{T}}{2T\sqrt{1-T}} ,
\end{equation}
which is the bimodal distribution of transmission eigenvalues in a lossless diffusive medium.

\par The prefactor $\frac{N_\cin}{N_{\rm e}}$ accounts for the normalization over the nonzero transmission eigenvalues,
$\textstyle\int_{0^+}^1\D{T}\,\rho(T)=1$.
Indeed, when $N_{\rm e}=N_\cout<N_\cin$, the $N_\cin\times N_\cin$ matrix $\herm{\matr{t}}\matr{t}$ contains $(N_\cin-N_{\rm e})$ additional zero eigenvalues, which are excluded from $\rho(T)$.
The full distribution, including these zero eigenvalues, is therefore
\begin{equation}\label{eq:bimodal-full}
\rho_{\rm full}(T) = \frac{N_{\rm e}}{N_\cin} \rho(T) + \left( 1 - \frac{N_{\rm e}}{N_\cin} \right) \delta(T) \:,
\end{equation}
and satisfies $\textstyle\int_0^1\D{T}\,\rho_{\rm full}(T)=1$.

\par Equation \eqref{eq:bimodal-normalized} therefore shows that the bimodal distribution follows from the matrix diffusion equation for a lossless diffusive medium of arbitrary structure; no quasi-one-dimensional assumption is required.
A similar proof of the bimodal distribution in arbitrary structure is given in Ref.\ \cite{Nazarov1994a}.
In the quasi-one-dimensional case, a different proof, still based on the matrix theory presented here, is given in Ref.\ \cite{GaspardD2025b}.

\subsection{Analytic profile of transmission eigenchannels}\label{app:itprofile-solution}%

\par The solution \eqref{eq:qn-param-lossless-2} directly provides the upper-right element of $\matr{Q}(\vect{r})$ entering Eq.\ \eqref{eq:itprofile-from-qn},
\begin{equation}\label{eq:q12-from-theta}
Q_{12}(\vect{r}) = \sqrt{\frac{T}{1-T}}  \sin\left[ \frac{\theta_\cin}{2}\bar{I}(\vect{r}) \right]  .
\end{equation}
Substituting Eq.\ \eqref{eq:q12-from-theta} into Eq.\ \eqref{eq:itprofile-from-qn} and using Eq.\ \eqref{eq:bimodal-normalized}, we obtain
\begin{equation}\label{eq:itprofile-from-re-sin}
I_{T}(\vect{r}) = \frac{2T}{\pi\bar{T}} \Re \sin\left[ \frac{\theta_\cin}{2}\bar{I}(\vect{r}) \right] .
\end{equation}
Using Eq.\ \eqref{eq:theta-input} and the identity $\Re\sin(x+\I y)=\cosh(y)\sin(x)$, this becomes
\begin{equation}\label{eq:itprofile-lossless}
I_{T}(\vect{r}) = \frac{2T}{\pi\bar{T}} \cosh\left[ \arccosh\left(\frac{1}{\sqrt{T}}\right) \bar{I}(\vect{r}) \right] \sin\left[ \frac{\pi}{2}\bar{I}(\vect{r}) \right] .
\end{equation}
This is the expression reported in the main text.

\subsection{Current density of transmission eigenchannels}\label{app:jtcurrent-solution}%

\par In this section, we show that, in a lossless medium, the average current density of transmission eigenchannels is directly proportional to the classical random-input current density.
Substituting the exact solution for $\vect{\matr{J}}(\vect{r})$ in a lossless medium, Eq.\ \eqref{eq:jn-from-jbar}, into Eq.\ \eqref{eq:jtcurrent-from-jn}, and using the bimodal law \eqref{eq:bimodal-normalized}, we obtain
\begin{equation}\label{eq:jtcurrent-from-jbar}
\vect{j}_{T}(\vect{r}) = \frac{T}{\bar{T}} \bar{\vect{j}}(\vect{r}) ,
\end{equation}
thus revealing the direct proportionality between the current density of transmission eigenchannels, $\vect{j}_{T}(\vect{r})$, and the current density of a random-input, $\bar{\vect{j}}(\vect{r})$.
This result \eqref{eq:jtcurrent-from-jbar} is nontrivial because it highlights that the intensity profile of the transmission eigenchannels, Eq.\ \eqref{eq:itprofile-lossless}, does not obey Fick's law---that is, the current \eqref{eq:jtcurrent-from-jbar} is not proportional to the intensity gradient: $\vect{j}_{T}(\vect{r})\neq -\frac{\ell}{d}\grad_{\vect{r}}I_{T}(\vect{r})$.

\subsection{Transmission eigenchannels are not diffusion eigenmodes}\label{app:irreducibility-to-diffusion}

\par Equation \eqref{eq:itprofile-lossless} also clarifies the relation between transmission eigenchannel profiles and classical diffusion.
In particular, the transmission eigenchannel profile does not, in general, correspond to an eigenmode of a diffusion equation, contrary to what has been suggested in Refs.\ \cite{Ojambati2016, vanTiggelenB2025}.

\par This can be seen most directly for the perfectly transmitting eigenchannel, $T=1$, for which Eq.\ \eqref{eq:itprofile-lossless} obeys
\begin{equation}\label{eq:i1-exact}
\lapl_{\vect{r}}I_1(\vect{r}) + \frac{\pi^2}{4} \abs{\grad_{\vect{r}}\bar{I}(\vect{r})}^2 I_1(\vect{r}) = 0 .
\end{equation}
In an arbitrary structure, $\abs{\grad_{\vect{r}}\bar{I}(\vect{r})}^2$ depends on position, so that Eq.\ \eqref{eq:i1-exact} is not an eigenvalue problem for the Laplacian.
It reduces to such a problem only when the gradient of the classical diffusive solution has constant magnitude, as occurs in transversely invariant structures.
In this case, we have $\bar{I}(\vect{r})=2\frac{L+z_0-x}{L+2z_0}$, and Eq.\ \eqref{eq:i1-exact} becomes
\begin{equation}\label{eq:i1-approx}
\lapl_{\vect{r}}\tilde{I}_1(\vect{r}) + \left(\frac{\pi}{L+2z_0}\right)^2 \tilde{I}_1(\vect{r}) = 0 .
\end{equation}
The tilde over $\tilde{I}_1(\vect{r})$ emphasizes the different between the solutions of Eq.\ \eqref{eq:i1-approx} and those of Eq.\ \eqref{eq:i1-exact}.
Equation \eqref{eq:i1-approx} explains why a diffusion-eigenmode interpretation can appear natural in transversely invariant systems, while it does not extend to more general structures.

\par As an explicit example, consider a three-dimensional disordered waveguide shaped as a spherical sector with inner radius $R_\cin$ and outer radius $R_\cout$, and inject the wave through the inner surface.
Neglecting the extrapolation length ($z_0=0$), the solution of Eq.\ \eqref{eq:ibar-diffusion} is
\begin{equation}\label{eq:ibar-spherical-sector}
\bar{I}(\vect{r}) = 2\frac{R_\cin(R_\cout - r)}{r(R_\cout-R_\cin)} ,
\end{equation}
where $r$ denotes the distance from the center of the apex of the spherical sector.
Substituting Eq.\ \eqref{eq:ibar-spherical-sector} into Eq.\ \eqref{eq:itprofile-lossless} for $T=1$ gives
\begin{equation}\label{eq:i1-exact-spherical-sector}
I_1(\vect{r}) \propto \sin\left(\pi\frac{R_\cin(R_\cout - r)}{r(R_\cout-R_\cin)}\right) .
\end{equation}

\par
\begin{figure}[ht]%
\includegraphics{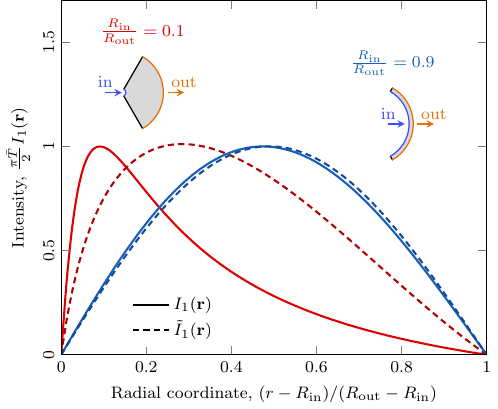}%
\caption{{\bf Open channel in a conical disordered waveguide.}
Intensity profile of the perfectly transmitting eigenchannel ($T=1$) in a three-dimensional conical waveguide for different ratios $R_\cin/R_\cout$.
Comparison between the profile given by Eq.\ \eqref{eq:i1-exact-spherical-sector}, obtained from Eq.\ \eqref{eq:itprofile-lossless} (continuous lines), and that given by Eq.\ \eqref{eq:i1-approx-spherical-sector}, obtained using the diffusion eigenmode ansatz \eqref{eq:i1-approx} (dashed lines).
The extrapolation distance is neglected ($z_0=0$).}%
\label{fig:conical-waveguide-v1}%
\end{figure}

 By contrast, treating the transmission eigenchannel as a diffusion eigenmode would lead to the diffusion equation \eqref{eq:i1-approx}, whose solution in the same structure and with the same boundary conditions ($\tilde{I}_1(R_{\cin})=\tilde{I}_1(R_{\cout})=0$) is
\begin{equation}\label{eq:i1-approx-spherical-sector}
\tilde{I}_1(\vect{r}) \propto \frac{\sin(\pi\frac{R_\cout - r}{R_\cout - R_\cin})}{\frac{2r}{R_\cin + R_\cout}} .
\end{equation}
The two predictions, Eqs.\ \eqref{eq:i1-exact-spherical-sector} and \eqref{eq:i1-approx-spherical-sector}, are compared in Fig.\ \ref{fig:conical-waveguide-v1} for different ratios $R_\cin/R_\cout$.

\par For $R_\cin\simeq R_\cout$, the two expressions coincide because the spherical sector approaches a transversely invariant waveguide.
In contrast, for $R_\cin\ll R_\cout$, the ansatz \eqref{eq:i1-approx-spherical-sector} breaks down and the two profiles differ substantially.
This provides an explicit example showing that the transmission eigenchannel profile is not, in general, an eigenmode of the diffusion equation.

\section{Transport on a sphere}\label{app:spherical-transport}%
In the presence of loss, the trajectory followed by $\matr{Q}(\vect{r})$ on the sphere $\matr{Q}(\vect{r})^2=\matr{1}_2$ is no longer confined to the great circle of the lossless solution in Eq.\ \eqref{eq:qn-param-lossless}, but becomes a more general trajectory involving two varying angular parameters.

\par To describe this trajectory, we first complete the matrix basis defined by $\matr{K}_2$ in Eq.\ \eqref{eq:def-k2-axis}.
We define
\begin{equation}\label{eq:def-k1-axis}
\matr{K}_1 = \frac{1}{2\I} [\matr{K}_2, \matr{Q}_\cout] = \frac{1}{\sqrt{1-T}} \begin{pmatrix}\I & \sqrt{T}\\ \frac{2-T}{\sqrt{T}} & -\I\end{pmatrix} ,
\end{equation}
and
\begin{equation}\label{eq:def-k3-axis}
\matr{K}_3 = \matr{Q}_\cout = \begin{pmatrix}1 & 0\\ \frac{-2\I}{\sqrt{T}} & -1\end{pmatrix} .
\end{equation}
The three basis matrices $\matr{K}_1,\matr{K}_2,\matr{K}_3$ obey the same algebra as the Pauli matrices,
\begin{equation}\label{eq:k-pauli-algebra}
\matr{K}_i \matr{K}_j = \matr{1}_2\delta_{ij} + \sum_{k=1}^{3} \I\varepsilon_{ijk}\matr{K}_k ,
\end{equation}
where $\varepsilon_{ijk}$ is the Levi-Civita symbol.

\par This algebra allows us to interpret conjugation by $\E^{-\I\alpha\matr{K}_i/2}$ as a rotation in the three-dimensional space spanned by the matrices $\matr{K}_i$.
We can therefore parametrize the general trajectory of $\matr{Q}(\vect{r})$ on the sphere as
\begin{equation}\label{eq:qn-kappa-param-1}
\matr{Q}(\vect{r}) = \E^{-\frac{\I}{2} \matr{K}_3 \phi(\vect{r})} \E^{-\frac{\I}{2} \matr{K}_2 \theta(\vect{r})} \matr{K}_3 \E^{\frac{\I}{2} \matr{K}_2 \theta(\vect{r})} \E^{\frac{\I}{2} \matr{K}_3 \phi(\vect{r})} .
\end{equation}
This extends the previous parametrization \eqref{eq:qn-param-lossless} to allow for a nonzero azimuthal angle $\phi(\vect{r})$.

\par Using the algebra \eqref{eq:k-pauli-algebra}, Eq.\ \eqref{eq:qn-kappa-param-1} can be written equivalently as
\begin{equation}\label{eq:qn-kappa-param-2}\begin{aligned}
\matr{Q}(\vect{r}) & = \cos\phi(\vect{r}) \sin\theta(\vect{r}) \matr{K}_1  \\
 & + \sin\phi(\vect{r}) \sin\theta(\vect{r}) \matr{K}_2  \\
 & + \cos\theta(\vect{r}) \matr{K}_3 ,
\end{aligned}\end{equation}
which makes the spherical nature of the parametrization explicit.
In particular, the upper-right element of $\matr{Q}(\vect{r})$, which provides the intensity profile of transmission eigenchannels according to Eq.\ \eqref{eq:itprofile-from-qn}, reads
\begin{equation}\label{eq:q12-kappa-param}
Q_{12}(\vect{r}) = \sqrt{\frac{T}{1-T}} \E^{-\I\phi(\vect{r})} \sin\theta(\vect{r}) ,
\end{equation}
thereby extending Eq.\ \eqref{eq:q12-from-theta} to nonzero $\phi(\vect{r})$.

\par
\begin{figure}[ht]%
\includegraphics{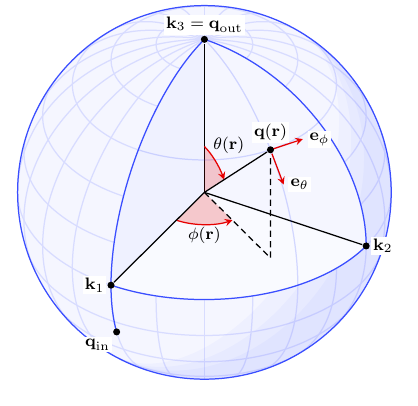}%
\caption{{\bf Spherical coordinate system used for $\vect{q}(\vect{r})$.}
Spherical manifold of $\vect{q}(\vect{r})$ according to the parametrization \eqref{eq:q-vector-kappa-param} and the constraint $\vect{q}(\vect{r})^2=1$.
The vectors $\vect{k}_1,\vect{k}_2,\vect{k}_3$ are the vector forms of the base matrices $\matr{K}_1,\matr{K}_2,\matr{K}_3$.}%
\label{fig:qn-manifold-detail-v3}%
\end{figure}

 The spherical geometry becomes more transparent in the corresponding vector representation.
We write
\begin{equation}
\matr{Q}(\vect{r}) = q_1(\vect{r}) \matr{K}_1 + q_2(\vect{r}) \matr{K}_2 + q_3(\vect{r}) \matr{K}_3 .
\end{equation}
The three components $(q_1,q_2,q_3)$ can be treated together as a vector,
\begin{equation}\label{eq:q-vector-from-matrix}
\vect{q}(\vect{r}) = \begin{pmatrix}q_1(\vect{r})\\ q_2(\vect{r})\\ q_3(\vect{r})\end{pmatrix}
 = \frac{1}{2} \begin{pmatrix}\Tr[\matr{K}_1\matr{Q}(\vect{r})]\\ \Tr[\matr{K}_2\matr{Q}(\vect{r})]\\ \Tr[\matr{K}_3\matr{Q}(\vect{r})]\end{pmatrix} ,
\end{equation}
which, according to Eq.\ \eqref{eq:qn-kappa-param-2}, reads
\begin{equation}\label{eq:q-vector-kappa-param}
\vect{q}(\vect{r}) = \begin{pmatrix}\cos\phi(\vect{r}) \sin\theta(\vect{r})\\ \sin\phi(\vect{r}) \sin\theta(\vect{r})\\ \cos\theta(\vect{r})\end{pmatrix} .
\end{equation}
The constraint $\matr{Q}(\vect{r})^2=\matr{1}_2$ translates into
\begin{equation}\label{eq:q-vector-sphere}
\vect{q}(\vect{r})^2 = 1,
\end{equation}
so that $\vect{q}(\vect{r})$ lies on a (complexified) unit sphere.
The parametrization \eqref{eq:q-vector-kappa-param} is represented graphically in Fig.\ \ref{fig:qn-manifold-detail-v3}.

\par The correspondence between the matrix and vector representations follows directly from the algebra \eqref{eq:k-pauli-algebra}.
Indeed, for two matrices $\matr{A}=\vect{a}\cdot\vect{\matr{K}}$ and $\matr{B}=\vect{b}\cdot\vect{\matr{K}}$, where $\vect{\matr{K}}=(\matr{K}_1,\matr{K}_2,\matr{K}_3)$, one has
\begin{equation}\label{eq:su2-so3-isomorphism}\begin{aligned}
\tfrac{1}{2} \Tr(\matr{A}\matr{B}) & = \vect{a}\cdot\vect{b} ,\\
\tfrac{1}{2\I} [\matr{A}, \matr{B}] & = (\vect{a}\times\vect{b})\cdot\vect{\matr{K}} ,\\
\tfrac{1}{2} \{\matr{A}, \matr{B}\} & = (\vect{a}\cdot\vect{b}) \matr{1}_2 ,\\
\E^{-\frac{\I}{2} \vect{\theta}\cdot\vect{\matr{K}}} \matr{Q} \E^{\frac{\I}{2} \vect{\theta}\cdot\vect{\matr{K}}} & = (\matr{R}(\vect{\theta})\,\vect{q})\cdot\vect{\matr{K}} ,
\end{aligned}\end{equation}
where $\matr{R}(\vect{\theta})$ is the standard three-dimensional rotation matrix
\begin{equation}
\matr{R}(\vect{\theta}) = \exp\left[ \begin{pmatrix}0 & -\theta_3 & \theta_2\\ \theta_3 & 0 & -\theta_1\\ -\theta_2 & \theta_1 & 0\end{pmatrix} \right] \:.
\end{equation}
These relations provide the matrix-vector correspondence used below.
In particular, the commutator of two $2\times2$ matrices is represented by the cross product of their corresponding vectors.

\par In the vector representation \eqref{eq:q-vector-from-matrix}, the matrix diffusion equation \eqref{eq:usadel-3rd-form}, which is equivalent to (omitting spatial dependence)
\begin{equation}\label{eq:usadel-commutator-form}
\left[ \lapl_{\vect{r}}\matr{Q} + \tfrac{d}{\ell\labso} \matr{\sigma}_3, \matr{Q} \right] = \matr{0},
\end{equation}
reads
\begin{equation}\label{eq:usadel-vector-alt}
\vect{\kappa}\times\vect{q} = \vect{0} ,\qquad \vect{\kappa} = \lapl_{\vect{r}}\vect{q} + \frac{d}{\ell\labso}\vect{\sigma}_3 ,
\end{equation}
where $\times$ denotes the vector product, and $\vect{\sigma}_3$ is given by
\begin{equation}\label{eq:sigma3-kappa-param}
\vect{\sigma}_3 = \frac{1}{2} \begin{pmatrix}\Tr(\matr{K}_1\matr{\sigma}_3)\\ \Tr(\matr{K}_2\matr{\sigma}_3)\\ \Tr(\matr{K}_3\matr{\sigma}_3)\end{pmatrix} = \begin{pmatrix}\frac{\I}{\sqrt{1-T}} \\ \frac{1}{\sqrt{1-T}} \\ 1\end{pmatrix} .
\end{equation}

\par Since $\vect{\kappa}\times\vect{q} = \vect{0}$ implies that $\vect{\kappa}$ is parallel to $\vect{q}$, while $\vect{q}_\theta=\partial_\theta\vect{q}$ and $\vect{q}_\phi=\partial_\phi\vect{q}$ are tangent to the sphere and therefore orthogonal to $\vect{q}$, Eq.\ \eqref{eq:usadel-vector-alt} reduces to two independent scalar equations,
\begin{equation}\label{eq:usadel-residual-alt}
\vect{\kappa}\cdot\vect{q}_\theta = 0 ,\qquad \vect{\kappa}\cdot\vect{q}_\phi = 0 .
\end{equation}
Substituting the parametrization \eqref{eq:q-vector-kappa-param} into Eq.\ \eqref{eq:usadel-residual-alt} then provides the parametrized form of the matrix diffusion equation,
\begin{equation}\label{eq:usadel-kappa-param}\begin{aligned}
& \lapl_{\vect{r}}\theta - \sin\theta\cos\theta(\grad_{\vect{r}}\phi)^2 = \tfrac{d}{\ell\labso} \left( \sin\theta - \tfrac{\I\E^{-\I\phi}\cos\theta}{\sqrt{1-T}} \right) ,\\
& \sin^2\theta\lapl_{\vect{r}}\phi + 2\sin\theta\cos\theta\grad_{\vect{r}}\theta\cdot\grad_{\vect{r}}\phi = -\tfrac{d}{\ell\labso} \tfrac{\E^{-\I\phi}}{\sqrt{1-T}} \sin\theta .
\end{aligned}\end{equation}

\par In the absence of absorption and free surfaces, as considered in the lossless case above, Eqs.\ \eqref{eq:usadel-kappa-param} admit the solution $\phi(\vect{r})=0$, corresponding to a trajectory confined to the great circle defined by $\matr{Q}_\cin$ and $\matr{Q}_\cout$.
The remaining equation then reduces to $\lapl_{\vect{r}}\theta = 0$, which is the scalar equation obtained in the lossless case discussed above.

\par Equation \eqref{eq:usadel-kappa-param} is accompanied by the input and output boundary conditions \eqref{eq:usadel-bdr-inout-2}, which translate in the parametrization \eqref{eq:qn-kappa-param-2} into
\begin{equation}\label{eq:usadel-kappa-bdr-inout}\begin{cases}
\theta(\vect{r}_\cin + z_0\vect{n}) = \theta_\cin ,\\
\phi(\vect{r}_\cin + z_0\vect{n}) = 0 ,
\end{cases} \quad \begin{cases}
\theta(\vect{r}_\cout + z_0\vect{n}) = 0 ,\\
\phi(\vect{r}_\cout + z_0\vect{n}) = 0 ,
\end{cases}\end{equation}
where $\theta_\cin$ is given by Eq.\ \eqref{eq:theta-input}.
In addition, the boundary condition at a free surface \eqref{eq:usadel-bdr-loss} becomes
\begin{equation}\label{eq:usadel-kappa-bdr-loss-1}\begin{cases}
\theta(\vect{r} + z_0\vect{n}) \rightarrow 0 ,\\
\phi(\vect{r} + z_0\vect{n}) \rightarrow -\I\infty ,
\end{cases}\end{equation}
under the constraint
\begin{equation}\label{eq:usadel-kappa-bdr-loss-2}
\theta(\vect{r} + z_0\vect{n}) \E^{\I\phi(\vect{r} + z_0\vect{n})} \rightarrow \frac{2\I}{\sqrt{1-T}} .
\end{equation}
\par It is interesting to note that Eqs.\ \eqref{eq:usadel-kappa-param} are closely analogous to the equations of motion of a three-dimensional spherical pendulum, upon replacing the position variable by time,
\begin{equation}\label{eq:pendulum-motion}\begin{aligned}
& \ddot{\theta} - \sin(\theta) \cos(\theta) \dot{\phi}^2 = \omega_0^2\vect{d}\cdot\vect{q}_\theta  ,\\
& \sin^2(\theta)\ddot{\phi} + 2\sin(\theta)\cos(\theta)\dot{\theta}\dot{\phi} = \omega_0^2\vect{d}\cdot\vect{q}_\phi ,
\end{aligned}\end{equation}
where the dots denote derivatives with respect to time, $\omega_0^2=g/l$ is the small-angle frequency, $g$ is the gravitational acceleration, and $l$ is the length of the pendulum.
The unit vector $\vect{d}$ defines the direction of gravity (the ``downward'' direction), which is oblique in the parametrization \eqref{eq:qn-kappa-param-2}.
The analogy between Eqs.\ \eqref{eq:usadel-kappa-param} and \eqref{eq:pendulum-motion} is obtained by identifying
\begin{equation}\label{eq:pendulum-gravity}
\omega_0^2\vect{d} = -\tfrac{d}{\ell\labso}\vect{\sigma}_3 ,
\end{equation}
where $\vect{\sigma}_3$ is given by Eq.\ \eqref{eq:sigma3-kappa-param}.

\par Finally, although useful for analytical calculations, Eq.\ \eqref{eq:usadel-kappa-param} is not suitable for numerical purposes for two reasons.
First, not only do the two base matrices $\matr{K}_1$ and $\matr{K}_2$ from Eqs.\ \eqref{eq:def-k1-axis} and \eqref{eq:def-k2-axis} diverge for $T\rightarrow 1$, i.e., 
\begin{equation}\label{eq:k1-k2-divergence}
\matr{K}_1 \xrightarrow{T\rightarrow 1} \begin{pmatrix}\I\infty & +\infty\\ +\infty & -\I\infty\end{pmatrix} ,\quad \matr{K}_2 \xrightarrow{T\rightarrow 1} \begin{pmatrix}+\infty & -\I\infty\\ -\I\infty & -\infty\end{pmatrix} ,
\end{equation}
but they also coalesce in this limit due to the fact that
\begin{equation}\label{eq:k1-k2-coalescence}
\matr{K}_1 - \I\matr{K}_2 = \begin{pmatrix}0 & 0\\ \frac{2\sqrt{1-T}}{\sqrt{T}} & 0\end{pmatrix} \xrightarrow{T\rightarrow 1} 0 ,
\end{equation}
thus hindering the convergence of iterative solvers such as Newton-Raphson.
Second, the free boundary condition $\phi\rightarrow-\I\infty$ in Eq.\ \eqref{eq:usadel-kappa-bdr-loss-1} is singular.
Numerical solvers for the matrix diffusion equation therefore require a more appropriate parametrization, which is discussed in Appendix \ref{app:numerical-usadel}.

\section{Numerical method for the matrix diffusion equation}\label{app:numerical-usadel}%

\par In this Appendix, we present the numerical method used to solve the matrix diffusion equation \eqref{eq:usadel-2nd-form}.
This method is implemented in the Usador program \cite{GaspardD2026-usador}, which is used to generate some of the figures in the main text.
As discussed above, the parametrization \eqref{eq:qn-kappa-param-2} becomes ill-conditioned as $T\rightarrow 1$, because the basis matrices $\matr{K}_1$ and $\matr{K}_2$ become singular and coalesce in this limit.
To circumvent this problem, we instead use the basis formed by the three Pauli matrices $(\matr{\sigma}_1,\matr{\sigma}_2,\matr{\sigma}_3)$, which is independent of $T$.
In this basis, we consider the parametrization
\begin{equation}\label{eq:qn-sigma-param-1}
\matr{Q}(\vect{r}) = \E^{-\frac{\I}{2}\matr{\sigma}_2\varphi(\vect{r})} \E^{-\frac{\I}{2}\matr{\sigma}_1\vartheta(\vect{r})} \matr{\sigma}_3 \E^{\frac{\I}{2}\matr{\sigma}_1\vartheta(\vect{r})} \E^{\frac{\I}{2}\matr{\sigma}_2\varphi(\vect{r})} ,
\end{equation}
or equivalently
\begin{equation}\label{eq:qn-sigma-param-2}\begin{aligned}
\matr{Q}(\vect{r}) & = \sin\varphi(\vect{r})\cos\vartheta(\vect{r}) \matr{\sigma}_1  \\
 & - \sin\vartheta(\vect{r}) \matr{\sigma}_2  \\
 & + \cos\varphi(\vect{r}) \cos\vartheta(\vect{r}) \matr{\sigma}_3 .
\end{aligned}\end{equation}
Like the parametrization \eqref{eq:qn-kappa-param-2}, this one is constructed so that the great circle connecting $\matr{Q}_\cin$ and $\matr{Q}_\cout$ coincides with a meridian of the parametrization, ensuring that the azimuthal coordinate $\varphi(\vect{r})$ remains constant in the absence of loss.
This can be achieved by placing the pole of the parametrization at any point on the great circle $\matr{Q}_\cin \to \matr{Q}_\cout$.
In the parametrization \eqref{eq:qn-kappa-param-2}, the pole was $\matr{Q}_\cout$; here in Eq.\ \eqref{eq:qn-sigma-param-2}, we choose instead the second Pauli matrix,
\begin{equation}\label{eq:sigma2-from-qin-qout}
\matr{\sigma}_2 = \frac{\matr{Q}_\cin-\matr{Q}_\cout}{\sqrt{(\matr{Q}_\cin-\matr{Q}_\cout)^2}} ,
\end{equation}
which belongs to the great circle (as does any linear combination of $\matr{Q}_\cin$ and $\matr{Q}_\cout$ upon normalization) but does not depend on $T$.
The parametrization \eqref{eq:qn-sigma-param-2} is therefore free from the singular behaviors \eqref{eq:k1-k2-divergence} and \eqref{eq:k1-k2-coalescence}, while maintaining a constant coordinate, $\varphi(\vect{r})$, in the absence of loss.
In the parametrization \eqref{eq:qn-sigma-param-2}, the upper-right element of $\matr{Q}(\vect{r})$ involved in Eq.\ \eqref{eq:itprofile-from-qn} reads
\begin{equation}\label{eq:q12-sigma-param}
Q_{12}(\vect{r}) = \sin\varphi(\vect{r})\cos\vartheta(\vect{r}) + \I\sin\vartheta(\vect{r}) ,
\end{equation}
which differs from Eq.\ \eqref{eq:q12-kappa-param} as a result of the change of parametrization.
In the parametrization \eqref{eq:qn-sigma-param-2}, the matrix diffusion equation \eqref{eq:usadel-2nd-form} becomes
\begin{equation}\label{eq:usadel-sigma-param}\begin{aligned}
& \lapl_{\vect{r}}\vartheta + \sin\vartheta\cos\vartheta(\grad_{\vect{r}}\varphi)^2 = \tfrac{d}{\ell\labso} \cos\varphi\sin\vartheta ,\\
& \cos\vartheta\lapl_{\vect{r}}\varphi - 2\sin\vartheta\grad_{\vect{r}}\vartheta\cdot\grad_{\vect{r}}\varphi = \tfrac{d}{\ell\labso} \sin\varphi .
\end{aligned}\end{equation}
Equation \eqref{eq:usadel-sigma-param} is accompanied by the input and output boundary conditions \eqref{eq:usadel-bdr-inout-2}, which read, according to Eq.\ \eqref{eq:qn-sigma-param-2},
\begin{equation}\label{eq:usadel-sigma-bdr-inout}\begin{cases}
\vartheta(\vect{r}_\cout + z_0\vect{n}) = -\vartheta(\vect{r}_\cin + z_0\vect{n}) = \frac{\pi}{2} + \I\arccosh(\frac{1}{\sqrt{T}}) ,\\
\varphi(\vect{r}_\cout + z_0\vect{n}) = \varphi(\vect{r}_\cin + z_0\vect{n}) = \frac{\pi}{2} - \I\arccosh(\frac{1}{\sqrt{1-T}}) ,
\end{cases}\end{equation}
and by the free-surface boundary condition \eqref{eq:usadel-bdr-loss}, which becomes
\begin{equation}\label{eq:usadel-sigma-bdr-loss}\begin{cases}
\vartheta(\vect{r} + z_0\vect{n}) = 0 ,\\
\varphi(\vect{r} + z_0\vect{n}) = 0 ,
\end{cases}\end{equation}
where $\vect{r}$ lies on the leaky edge.
The boundary condition \eqref{eq:usadel-sigma-bdr-loss} is thus nonsingular, in contrast to Eq.\ \eqref{eq:usadel-kappa-bdr-loss-1}.

\par Equation \eqref{eq:usadel-sigma-param} is solved numerically by the Usador program \cite{GaspardD2026-usador} on a two-dimensional square lattice using the Newton-Raphson iterative algorithm.
An appropriate initial guess for this algorithm is
\begin{equation}\label{eq:usador-initial-guess}
\vartheta^{(0)}(\vect{r}) = 0, \quad
\varphi^{(0)}(\vect{r}) = \tfrac{\pi}{2} - \I\arccosh(\tfrac{1}{\sqrt{1-T}}) ,
\end{equation}
which has the particularity of falling midway between the input and output boundary conditions of Eq.\ \eqref{eq:usadel-sigma-bdr-inout}.

\par Finally, in the absence of absorption and free surfaces, the constant-$\varphi$ solution of Eq.\ \eqref{eq:usadel-sigma-param} reduces to
\begin{equation}\label{eq:usadel-sigma-lossless}
\lapl_{\vect{r}}\vartheta(\vect{r}) = 0 ,
\end{equation}
similarly to Eq.\ \eqref{eq:usadel-kappa-lossless}.
Therefore, the results of Appendix \ref{app:lossless-solution} could in principle be retrieved using the parametrization \eqref{eq:qn-sigma-param-2}.
However, the calculation would be more tedious than with \eqref{eq:qn-kappa-param-2} because the boundary conditions \eqref{eq:usadel-sigma-bdr-inout} and the expressions of $Q_{12}(\vect{r})$ and $\vect{J}_{12}(\vect{r})$ are more complex.
In addition, several useful expressions with particularly transparent geometrical interpretations would be altered in the parametrization \eqref{eq:qn-sigma-param-2}, in particular Eqs.\ \eqref{eq:def-k2-axis}, \eqref{eq:jn-from-grad-theta}, and \eqref{eq:ibar-from-theta-alt}.

\section{Additional results}\label{app:additional-results}
In this Appendix, we present two additional results supporting the findings of the main text: on the one hand, the transmission eigenchannels in a slab at different sizes of the illumination region, and on the other hand, the numerical verification of the transmission gap in a disordered two-duct waveguide.

\subsection{Focused waves in a disordered slab}\label{app:slab-detail}
\begin{figure*}[p]%
\includegraphics{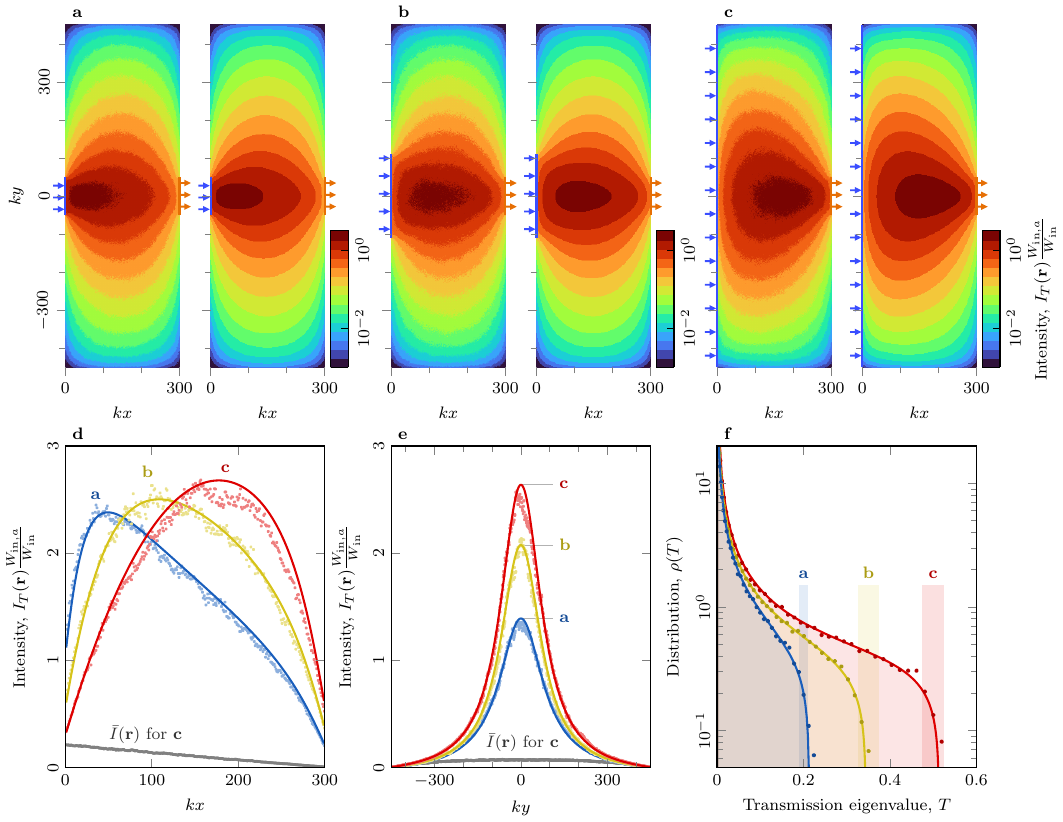}%
\caption{{\bf Effect of the illumination size on focused waves through a disordered slab.}
{\bf a}, {\bf b}, {\bf c}, Intensity profiles of focused waves through a two-dimensional disordered slab for different sizes of the illumination region.
The scattering thickness is $L/\ltran=20$, the slab width $W=3L$, and the target region $W_\cout=W/9$.
Left panels are wave-equation simulations averaged over $N_{\rm s}>300$ realizations of the disorder, and right panels predictions of Eqs.\ \eqref{eq:usadel-2nd-form} and \eqref{eq:itprofile-from-qn} for $\labso^{-1}\rightarrow 0$ with the boundary conditions \eqref{eq:usadel-bdr-inout-1} in controlled regions (arrows) and \eqref{eq:usadel-bdr-loss} in free regions.
The intensities $I_{T}(\vect{r})$ are multiplied by $W_{\cin,a}/W_{\cin}$, where $W_{\cin,a}$ is the illumination size in panel {\bf a}, in order to compare similar injection powers.
{\bf a}, $W_\cin=W_\cout$, and $T=20\%$. In the left panel, $N_{\rm s}=620$.
{\bf b}, $W_\cin=\frac{11}{5}W_\cout$, and $T=35\%$. In the left panel, $N_{\rm s}=381$.
{\bf c}, $W_\cin=9W_\cout$, and $T=50\%$. In the left panel, $N_{\rm s}=461$.
{\bf d}, {\bf e}, Cross sections of focused waves intensities corresponding to {\bf a}, {\bf b}, {\bf c}.
Dots are wave-equation simulations (averaged spatially over a $2\lambda$ window), and lines are predictions of Eq.\ \eqref{eq:usadel-2nd-form}.
The gray dots are simulations of $\bar{I}(\vect{r})$ for $W_\cin=9W_\cout$.
{\bf d}, Horizontal cross section at $ky=0$.
{\bf e}, Vertical cross section at $kx=200$.
{\bf f}, Distribution of transmission eigenvalues $\rho(T)$ corresponding to {\bf a}, {\bf b}, {\bf c}.
Dots are wave-equation simulations, and lines predictions of Eqs.\ \eqref{eq:usadel-2nd-form} and \eqref{eq:tspectrum-from-jn}.
Rectangles represent the intervals of $T$ over which transmission eigenchannels are averaged in wave-equation simulations.}%
\label{fig:slab-detail-v2}%
\end{figure*}

In Fig.\ \ref{fig:slab-detail-v2}a--e, we present the intensity profile of open channels in a disordered slab for different sizes of the illuminations region.
As the illumination region expands, the intensity peak moves toward the targeted region, as can be clearly seen in the horizontal cross section of Fig.\ \ref{fig:slab-detail-v2}d.
This phenomenon, observed in simulations based on the wave equation (left panels of Fig.\ \ref{fig:slab-detail-v2}a--c), is predicted by the matrix diffusion equation \eqref{eq:usadel-2nd-form} for $\labso^{-1}\rightarrow 0$ (right panels of Fig.\ \ref{fig:slab-detail-v2}a--c).

\par The vertical cross section in Fig.\ \ref{fig:slab-detail-v2}e depicts an exponential decay in the transverse direction far from the intensity peak ($\abs{y}\rightarrow\infty$).
We have checked that this decay matches the behavior predicted by the classical diffusion equation $\lapl_{\vect{r}}I_{T}(\vect{r}) = 0$, that is
\begin{equation}\label{eq:itprofile-slab-asym}
I_{T}(\vect{r}) \overset{\abs{y}\rightarrow\infty}{\propto} \exp\left(-\frac{\pi\abs{y}}{L+2z_0}\right) \sin\left(\pi\frac{x+z_0}{L+2z_0}\right) ,
\end{equation}
where $L$ is the horizontal thickness of the slab.
This behavior can be explained from the matrix diffusion equation \eqref{eq:usadel-2nd-form}.
Indeed, in a region far enough from the illumination and dominated by losses, we expect that the off-diagonal elements of $\matr{Q}(\vect{r})$ are small, i.e., that $\abs{Q_{12}(\vect{r})}\ll 1$ and $\abs{Q_{21}(\vect{r})}\ll 1$, because of the influence of the boundary condition \eqref{eq:usadel-bdr-loss}.
In this case, at first order in $Q_{12}(\vect{r})$ and $Q_{21}(\vect{r})$, Eq.\ \eqref{eq:usadel-2nd-form} reduces to linear diffusion problems:
\begin{equation}
\lapl_{\vect{r}}Q_{12}(\vect{r}) = 0 ,\qquad \lapl_{\vect{r}}Q_{21}(\vect{r}) = 0 .
\end{equation}
Since $I_{T}(\vect{r})\propto\Re Q_{12}(\vect{r})$, we deduce that $I_{T}(\vect{r})$ should obey classical diffusion, hence the behavior \eqref{eq:itprofile-slab-asym}.
Physically, this means that in a region far from the illumination and dominated by losses, the wave loses memory of its initial shaped wavefront and classical diffusion applies.

\par In Fig.\ \ref{fig:slab-detail-v2}f, we also verify that the transmission eigenvalue distribution $\rho(T)$ predicted by Eqs.\ \eqref{eq:usadel-2nd-form} and \eqref{eq:tspectrum-from-jn} (solid lines) indeed coincides with the distribution obtained by simulating the wave equation (dots) for the three different sizes of the illumination region.

\subsection{Transmission gap in a two-duct disordered waveguide}\label{app:circuit-check}
\begin{figure*}[p]%
\includegraphics{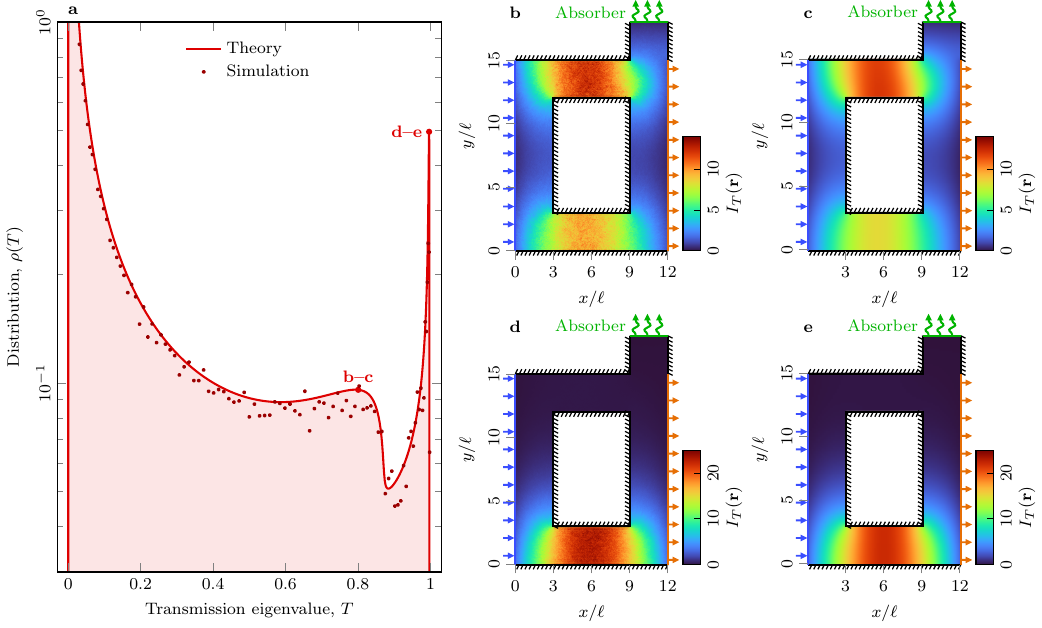}%
\caption{{\bf Transmission gap in a two-duct disordered waveguide with an absorber.}
Comparison between predictions of Eq.\ \eqref{eq:usadel-2nd-form} and simulations based on the wave equation.
All wave-equation simulations assume 333 input and output modes, the mean free path $\ell=12.8\,\lambda$, and are averaged over 500 realizations of the disorder.
{\bf a}, Distribution of transmission eigenvalues in the two-duct waveguide with an absorber.
Dots are the wave-equation simulations, and the solid line is the prediction of Eq.\ \eqref{eq:usadel-2nd-form}.
{\bf b}, Intensity profile computed from the wave equation for the transmission eigenchannel at $T=80\%$.
{\bf c}, Intensity profile predicted by Eq.\ \eqref{eq:usadel-2nd-form} for $T=80\%$.
{\bf d}, Intensity profile computed from the wave equation for $T=99.6\%$.
{\bf e}, Intensity profile predicted by Eq.\ \eqref{eq:usadel-2nd-form} for $T=99.6\%$.}%
\label{fig:circuit-check-v3}%
\end{figure*}

In Fig.\ \ref{fig:circuit-check-v3}a, we validate the gap in the transmission eigenvalue distribution predicted for a two-duct disordered waveguide with a localized absorber by the matrix diffusion equation \eqref{eq:usadel-2nd-form} (solid line) using numerical simulations based on the wave equation (dots).
In Fig.\ \ref{fig:circuit-check-v3}b--e, we also confirm the symmetry breaking phenomenon of transmission eigenchannels across the gap, using simulations of the wave equation.

\end{document}